\documentclass[aps,groupedaddress,amsmath,showkeys,eqsecnum]{revtex4}
\usepackage{bm}
\usepackage{graphicx,amssymb}
\usepackage{bm}
\usepackage{amsmath}
 \usepackage{color}

\newcommand{\kk}{{\text{th}}}
\newcommand{\beq}{\begin{equation}}
\newcommand{\eeq}{\end{equation}}
\newcommand{\beqa}{\begin{eqnarray}}
\newcommand{\eeqa}{\end{eqnarray}}
\newcommand{\nn}{\nonumber\\}
\newcommand{\TT}{\widetilde{T}}
\newcommand{\zetat}{\widetilde{\zeta}_{1}}
\newcommand{\dd}{\text{d}}

\begin{document}

\title{Impurity in a granular gas under nonlinear Couette flow}
\author{Francisco Vega Reyes}
\email{fvega@unex.es}
\author{Vicente Garz\'o}
\email{vicenteg@unex.es}
\homepage{http://www.unex.es/eweb/fisteor/vicente/}
\author{Andr\'es Santos}
\email{andres@unex.es}
\homepage{http://www.unex.es/eweb/fisteor/andres/}
\affiliation{Departamento de F\'isica, Universidad de
Extremadura, E--06071 Badajoz, Spain}

\date{\today}
\begin{abstract}
{We study in this work} the transport properties of an impurity
immersed in a granular gas under
 stationary nonlinear  Couette flow. The starting point is a
 kinetic model for low-density granular mixtures recently proposed by the
authors [{Vega Reyes F et al. 2007 {\em Phys.\ Rev.\ E} {\bf 75}
061306}]. Two routes have been considered. First,  a {hydrodynamic}
or normal solution is found by exploiting a formal mapping between
the kinetic equations for the gas particles and for the impurity. We
show that the transport properties of the impurity are characterized
by the ratio between the temperatures of the impurity and gas
particles and by five generalized transport coefficients: three
related to the momentum flux (a nonlinear shear viscosity and two
normal stress differences) and two related to the heat flux (a
nonlinear thermal conductivity and a cross coefficient measuring a
component of the heat flux orthogonal to the thermal gradient).
{Second, by means of a Monte Carlo simulation method we numerically
solve the kinetic equations and show that our hydrodynamic solution
is valid in the bulk of the fluid when realistic boundary conditions
are used. Furthermore, the hydrodynamic solution applies to
arbitrarily (inside the continuum regime) large values of the shear
rate, of the inelasticity, and of the rest of parameters of the
system.} {Preliminary simulation results of the true Boltzmann
description show the reliability of the nonlinear hydrodynamic
solution of the kinetic model.} This shows again the validity of a
hydrodynamic description for granular flows, even under extreme
conditions, beyond the Navier--Stokes domain.

\end{abstract}
 \maketitle

\section{Introduction\label{sec1}}
The understanding of transport processes occurring in granular
mixtures is still challenging. In the low- and moderate-density
regimes the Boltzmann and Enskog equations, suitably adapted to
account for inelastic collisions, have proven to provide an adequate
framework for the study of granular flows \cite{G03,BP04}. In
particular, if the spatial gradients present in the system are weak,
the Navier--Stokes (NS) constitutive equations for the fluxes of
mass, momentum, and energy have been derived (with explicit
expressions for the transport coefficients) for the model of
inelastic hard spheres characterized by constant coefficients of
normal restitution $\alpha_{ij}$. Most of the early derivations were
restricted to the quasielastic limit ($\alpha_{ij}\approx 1$), thus
assuming an expansion around Maxwellians at the \emph{same}
temperature \cite{JM87,JM89,Z95,AW98,WA99,SGNT06}. However, the
 nonequipartition of energy becomes significant beyond the
 quasi-elastic limit, as confirmed by kinetic theory \cite{MP99,GD99b,BT02,G06}, computer
 simulations \cite{BT02,MG02,DHGD02,PMP02,PCMP03,KT03,WJM03,BRM05,BRM06,SUKSS06}, and real experiments
 \cite{SUKSS06,WP02,FM02}.
{A more realistic derivation of the NS transport coefficients
\cite{GD02,GMD06,GM07,GDH07} requires taking into account the
nonequipartition of energy and applies for finite dissipation.} The
accuracy of this latter approach has
 been confirmed by computer simulations in the cases of the diffusion
  \cite{BRCG00,GM04} and  shear viscosity
 \cite{MG03,GM03} coefficients.
{On the other hand, the practical applicability of the NS equations
is limited to small  spatial gradients, while many steady granular
flows do not fulfill in general this condition, due to the coupling
between inelasticity and gradients \cite{G03,SGD04}.}

\begin{figure}
\includegraphics[width=.5\columnwidth]{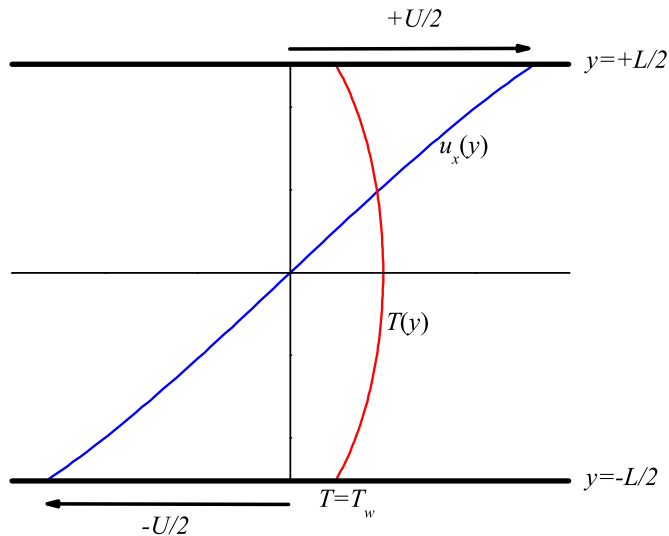}
\caption{(Color online) Sketch of a planar Couette gas flow. The gas
is enclosed between two infinite parallel walls located at $y=\pm
L/2$, moving along the $x$-direction with velocities $\pm U/2$, and
kept at the temperature $T_w$.}
\label{fig1}
\end{figure}

 The physical situation we study in this work corresponds to a gas of inelastic hard
spheres enclosed between two parallel walls at $y=\pm L/2$ moving
with velocities $\pm U/2$ along the $x$-axis and kept, in general,
at different  temperatures {$T_{\pm}$}
\cite{RC88,HJR88,L96,B97,TTMGSD01,LMG07}. In the base steady state
the flow velocity is along the $x$-axis and the hydrodynamic fields
depend on the $y$ variable only (planar Couette flow). This
macroscopic state is characterized by a combined momentum and heat
transport described by the pressure tensor $P_{ij}(y)$ and the heat
flux $\mathbf{q}(y)$, respectively. {A sketch of the geometry of the
steady planar Couette flow for the symmetric choice
{$T_{+}=T_{-}=T_{w}$} is given in Fig.\ \ref{fig1}.}

Since granular matter is generally present in polydisperse form, the
study of the Couette flow in the case of a granular mixture is an
interesting problem  from a fundamental and practical point of view.
Needless to say, the general problem is quite intricate since, not
only the number of parameters (masses, sizes, composition, and
coefficients of restitution)  but also the number of transport
coefficients  are  higher than in the monocomponent case. As a first
step and to gain some insight into the general problem, in this
paper we consider the tracer limit  case, namely a binary mixture
where the mole fraction of one of the components (tracer species,
denoted by the label 1) is much smaller than that of the other
component (excess species, denoted by the label 2). In this tracer
limit, the state of the excess species is unaffected by the presence
of the tracer particles and so its velocity distribution function
$f_2$ obeys a closed Boltzmann equation {in the low-density regime}.
In addition, the mutual collisions among the tracer particles can be
neglected versus the tracer-gas  collisions, so that the tracer
velocity distribution function $f_1$ obeys a linear (inelastic)
Boltzmann--Lorentz equation. This problem is formally equivalent to
that of an impurity or intruder immersed in a granular gas, and this
will be the terminology used in this paper. Since the impurity
particle is assumed to be mechanically different from the gas
particles, the dimensionless parameters characterizing the mixture
are the coefficients of restitution $\alpha_{12}$ and $\alpha_{22}$,
the mass ratio $m_1/m_2$, and the size ratio $\sigma_1/\sigma_2$.

Unfortunately, the complexity of the nonlinear Couette flow makes
its treatment at the level of the Boltzmann equation practically
unattainable, even in the monocomponent case. Thus,  here we will
consider a model kinetic equation recently proposed for granular
mixtures \cite{VGS07}. In the tracer limit, this kinetic model
reduces to the same closed kinetic equation for the excess species
as considered in Ref.\ \cite{TTMGSD01} plus a
Boltzmann--Lorentz-like  kinetic equation for the impurity particle.
{The kinetic equation for $f_2$ admits an \emph{exact} solution for
the steady planar Couette flow \cite{TTMGSD01}.} Exploiting the
formal analogy between the kinetic equations for $f_1$ and $f_2$, we
find in this paper an exact solution for $f_1$. This solution allows
us to obtain the most relevant velocity moments of $f_1$, which are
directly related to the momentum and heat fluxes associated with the
impurity. In particular, as expected, the impurity temperature
clearly differs from the granular temperature of the gas particles,
showing again the breakdown of the energy equipartition in
nonequilibrium states.

The exact solution found here qualifies as a {``normal''} or
hydrodynamic solution since $f_1$ and $f_2$ depend on space only
through an explicit functional dependence on the hydrodynamic
fields. This hydrodynamic description applies even at strong
dissipation (i.e., beyond the quasi-elastic limit) and strong
inhomogeneity (i.e., beyond the NS domain), as long as the densest
regions of the system remain sufficiently dilute and the molecular
chaos assumption holds. This provides a counter-example against the
speculation that a hydrodynamic description for granular flows is
limited to weak dissipation and/or weak inhomogeneities. In order to
assess the reliability of this hydrodynamic solution, we have also
solved the model kinetic equation by means of Monte Carlo
simulations with Couette-flow boundary conditions. Comparison with
the hydrodynamic solution shows that the latter is not a
mathematical artifact but applies in the bulk region of the system,
where boundary effects are negligible. This agreement between theory
and simulations holds for system sizes $L$ as small as a few mean
free paths.

In order to gain some insight into the expected hydrodynamic fields
in the Couette problem, let us consider first a monocomponent
granular gas. In this case, the {exact energy and momentum balance
equations yield} \beq \frac{2}{dn}\left(P_{xy}\frac{\partial
u_x}{\partial y}+\frac{\partial q_y}{\partial y}\right)=-\zeta T,
\label{1} \eeq \beq \frac{\partial P_{xy}}{\partial y}=0, \label{2}
\eeq \beq \frac{\partial P_{yy}}{\partial y}=0, \label{2bis} \eeq
where $d=2$ and 3 for hard disks and spheres, respectively, $n$ is
the number density, and $\zeta$ is the cooling rate due to the
inelastic character of collisions. By dimensional analysis {in the
dilute limit}, $\zeta=\nu\zeta^*(\alpha)$, where $\nu\propto n
T^{1/2}$ is an effective collision frequency for hard spheres.
{Equations \eqref{1}--\eqref{2bis}} do not constitute a closed set
of equations for the hydrodynamic fields $n(y)$, $T(y)$, and
$u_x(y)$, unless the constitutive equations expressing the fluxes as
functionals of the hydrodynamic fields are known. For illustration,
let us assume for the moment that the hydrodynamic gradients are
small enough as to justify the use of the NS constitutive equations.
Due to the geometry of the problem, at NS order we have
$P_{xx}=P_{yy}=P_{zz}=p$ \cite{BDKS98,GD99a}, from which, with
\eqref{2bis}, it immediately follows that the hydrostatic pressure
$p=nT$ is constant, i.e., \beq p=\text{const}. \label{5} \eeq
{Moreover, the NS constitutive equations imply that $q_x=q_z=0$ and}
\beq P_{xy}=-\eta \frac{\partial u_x}{\partial y}, \quad
q_y=-\kappa\frac{\partial T}{\partial y}-\mu \frac{\partial
n}{\partial y}, \label{4} \eeq where $\eta=(p/\nu)\eta^*(\alpha)$ is
the shear viscosity, $\kappa=(p/m\nu)\kappa^*(\alpha)$ is the
thermal conductivity ($m$ being the mass of a particle), and
$\mu=(T^2/m\nu)\mu^*(\alpha)$ is a transport coefficient absent in
the elastic case ($\alpha=1$). The explicit form of  the
dimensionless functions {$\zeta^*(\alpha)$, $\eta^*(\alpha)$,
$\kappa^*(\alpha)$, and $\mu^*(\alpha)$  is known
\cite{BDKS98,GD99a}.} Insertion of Eqs.\ \eqref{5} and \eqref{4}
into Eqs.\ \eqref{1} and \eqref{2} yield \beq a\equiv
\frac{1}{\nu}\frac{\partial u_x}{\partial y}=\text{const}, \label{6}
\eeq \beq \frac{1}{2m}\left(\frac{1}{\nu}\frac{\partial}{\partial
y}\right)^2 T=-\gamma =\text{const}. \label{7} \eeq Therefore,
according to the NS description, the local shear rate $\partial
u_x/\partial y$ scaled with the local collision frequency
$\nu\propto p/T^{1/2}$ is  a constant, and the temperature profile
is such that $(\nu^{-1}\partial_y)^2T$ is a constant that depends on
the reduced shear rate $a$ and the coefficient of restitution
$\alpha$. {The set of NS base steady states in the system have been
analytically solved in a recent work \cite{VU07}.}

\begin{figure}
\includegraphics[width=.5\columnwidth]{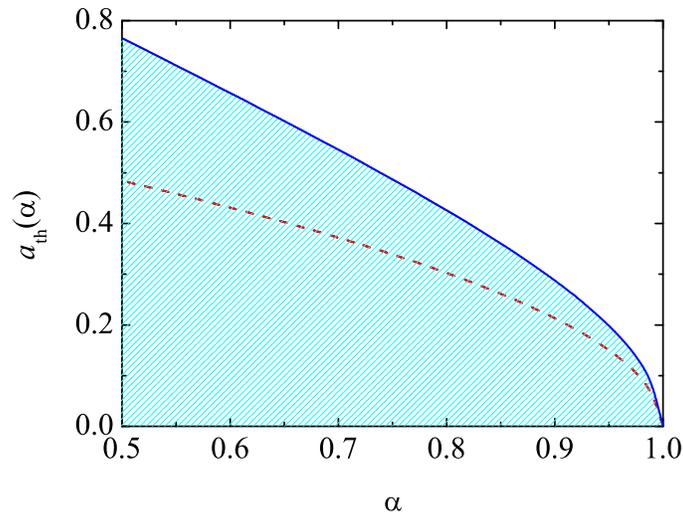}
\caption{(Color online) Plot of the threshold value of the reduced
shear rate, $a_\text{th}(\alpha)$ for a three-dimensional granular
gas in the planar Couette flow. The dashed line is the result
$a_\text{th}^\text{NS}(\alpha)=\sqrt{d\zeta^*(\alpha)/2\eta^*(\alpha)}$
obtained from the NS equations, while the solid line is the
prediction
$a_\text{th}(\alpha)=\sqrt{d\zeta^*(\alpha)/2}[1+\zeta^*(\alpha)]$
from an exact solution of a kinetic model of the Boltzmann equation
\protect\cite{TTMGSD01}. The separation between both curves is a
measure of the limitations of the NS description.}
\label{fig2}
\end{figure}

As said before, due to inelasticity these steady states do not have
small spatial gradients (except for $\alpha\approx 1$ \cite{VU07})
and thus a kinetic description beyond the NS domain is in general
required in order to properly describe granular Couette flows.
Specifically, this is even more necessary {if} $\gamma\ge0$ (and
this happens when the viscous heating dominates over collisional
cooling \cite{SGD04}), since in this case the Knudsen number is
always greater than the one for $\gamma<0$ \cite{VU07}. Such a
description of the Couette flow beyond the NS domain was carried out
in Ref.\ \cite{TTMGSD01} for a monocomponent granular gas with
$\gamma\ge0$. Interestingly enough, this solution shares with the NS
description {the structure of the hydrodynamic profiles \eqref{5},
\eqref{6}, and \eqref{7}. However, in the constitutive equations,
the transport coefficients and the parameter $\gamma$ are
\emph{nonlinear} functions of the shear rate $a$ \cite{TTMGSD01}.}
At the same time, the solution is also able to capture normal stress
differences ($P_{xx}\neq P_{yy}\neq P_{zz}$) and {the component of}
the heat flux along the flow direction ($q_x\neq 0$), which are all
nonlinear effects \cite{TTMGSD01}. All theoretical results in Ref.\
\cite{TTMGSD01} compare well with Monte Carlo simulations of the
Boltzmann equation, showing the reliability of the kinetic model
beyond the quasi-elastic limit. As an illustrative example of the
necessity of a nonlinear description, we briefly analyze the case
$\gamma=0$, which occurs for a threshold value of $a$ that in the NS
description is
$a_\text{th}^\text{NS}(\alpha)=\sqrt{d\zeta^*(\alpha)/2\eta^*(\alpha)}$
and in the nonlinear Couette flow is
$a_\text{th}(\alpha)=\sqrt{d\zeta^*(\alpha)/2}[1+\zeta^*(\alpha)]$
\cite{TTMGSD01}. We show in Fig.\ \ref{fig2} the disagreement
between both values, which becomes very apparent for values far from
the quasielastic limit. {As shown in Ref.\ \cite{TTMGSD01}, the
nonlinear prediction $a_\text{th}(\alpha)$ agrees very well with
Monte Carlo simulations of the Boltzmann equation.}

{We propose in this work a theoretical solution of the nonlinear
hydrodynamic profiles for the impurity that {exhibits} absence of
mutual diffusion (i.e, flow velocities are equal for impurity and
excess components). Furthermore, this solution for the impurity also
obeys equations of the form \eqref{5}, \eqref{6}, and \eqref{7}. We
will use a numerical solution of the kinetic equation {by a Monte
Carlo method} in order to show that the theoretical solution we
propose matches the hydrodynamic profiles and transport coefficients
that result from the kinetic equation. Furthermore, with the use of
the numerical solution we show that the hypotheses, notably the
absence of mutual diffusion, used in order to find our hydrodynamic
solution are actually always true in a wide range of system
parameters (including shear rate and inelasticity).} {In addition,
we present preliminary Monte Carlo simulations  of the Boltzmann
equations which confirm the hydrodynamic profiles predicted by the
nonlinear hydrodynamic solution of the kinetic model.}

This paper is organized as follows. The kinetic model for the
mixture is described in Sec.\ \ref{sec2}. Then, the physical problem
we are interested in is introduced. Section \ref{sec3} presents the
exact hydrodynamic solution to the kinetic equations for $f_1$ and
$f_2$, with explicit expressions for the heat and momentum fluxes of
both species. The simulation method is described in Sec.\ \ref{sec4}
and  comparisons between the theoretical predictions and the
simulation results is carried out in Sec.\ \ref{sec5}. Finally, the
results are summarized and discussed in Sec.\ \ref{sec6}.

\section{Kinetic model for granular mixtures \label{sec2}}
Let us consider a  mixture composed by smooth inelastic disks
($d=2$) or spheres ($d=3$) of masses $m_{i}$ and diameters $\sigma
_{i}$, the inelasticity of collisions between a sphere of species
$i$ and a sphere of species $j$  being characterized by a constant
coefficient of restitution $0<\alpha _{ij}\leq 1$. {We will focus on
the dilute limit, i.e., the mean free path of the particles is much
larger than  their sizes}. The relevant hydrodynamic fields are the
number densities $n_{i}$, the flow velocity $ {\bf u}$, and the
temperature $T$. They are defined in terms of moments of the
velocity distribution functions $f_{i}({\bf r},{\bf v};t)$ as
\begin{equation}
n_{i}=\int \dd{\bf v}f_{i}({\bf v}),
\label{2.4.1}
\end{equation}
\begin{equation}
 \rho {\bf
u}=\sum_{i}m_{i}n_i \mathbf{u}_i=\sum_{i}m_{i} \int \dd {\bf
v}\,{\bf v}f_{i}({\bf v}),
\label{2.4}
\end{equation}
\begin{equation}
nT=p=\sum_{i} n_i T_i=\sum_{i}\frac{m_{i}}{d}\int \dd{\bf
v}\,V^{2}f_{i}({\bf v}), \label{2.5}
\end{equation}
where ${\bf V}={\bf v}-{\bf u}$ is the peculiar velocity,
$n=\sum_{i} n_{i}$ is the total number density,
$\rho=\sum_{i}\rho_i=\sum_{i} m_{i}n_{i}$ is the total mass density,
and $p$ is the pressure. Furthermore, the second equality of Eq.\
\eqref{2.4} and the third equality of Eq.\ (\ref{2.5}) define the
flow velocity $\mathbf{u}_i$ and the partial kinetic temperature
$T_i$ for each species, respectively. In addition, in the dilute
limit the pressure tensor $\mathsf{P}_i$ and the heat flux
$\mathbf{q}_i$ associated with species $i$ are given by
\beq
\mathsf{P}_i=m_i\int \dd\mathbf{v}\, \mathbf{V}\mathbf{V}
f_i(\mathbf{v}),\quad \mathbf{q}_i= \frac{m_i}{2}\int
\dd\mathbf{v}\, V^2\mathbf{V} f_i(\mathbf{v}).
\label{III.16}
\eeq

In the low-density regime the distribution functions $f_{i}$ {obey a
set of coupled nonlinear Boltzmann equations \cite{GD02}:
\begin{equation}
\left(\partial_t+\mathbf{v}\cdot\nabla\right)f_i=\sum_{j} J_{ij}[
{\bf v}|f_i,f_j], \label{n5.1}
\end{equation}}
{where  $J_{ij}[ {\bf v}|f_i,f_j]$ denotes the inelastic Boltzmann
 operator that gives the rate of change of $f_i$ due to collisions
 with particles of species $j$. It is given by
\begin{eqnarray}
J_{ij}\left[ {\bf v}_{1}|f_{i},f_{j}\right] &=&\sigma
_{ij}^{d-1}\int \dd{\bf v} _{2}\int \dd\widehat{\boldsymbol {\sigma
}}\,\Theta (\widehat{{\boldsymbol {\sigma }}} \cdot {\bf
g}_{12})(\widehat{\boldsymbol {\sigma }}\cdot {\bf g}_{12})
\nonumber
\\
&&\times \left[ \alpha _{ij}^{-2}f_{i}({\bf r},{\bf v}_{1}^{\prime
},t)f_{j}( {\bf r},{\bf v}_{2}^{\prime },t)-f_{i}({\bf r},{\bf v}
_{1},t)f_{j}({\bf r}, {\bf v}_{2},t)\right] .
\label{2.2}
\end{eqnarray}}
{In Eq.\ (\ref{2.2}), $d$ is the dimensionality of the system,
$\sigma _{ij}=\left( \sigma _{i}+\sigma _{j}\right) /2$,
$\widehat{\boldsymbol {\sigma}}$ is a unit vector along the line of
centers, $\Theta $ is the Heaviside step function, and ${\bf
g}_{12}={\bf v}_{1}-{\bf v}_{2}$ is the relative velocity. The
primes on the velocities denote the initial values $\{{\bf
v}_{1}^{\prime }, {\bf v}_{2}^{\prime }\}$ that lead to $\{{\bf
v}_{1},{\bf v}_{2}\}$ following a binary (restituting) collision:
\begin{subequations}
\begin{equation}
{\bf v}_{1}^{\prime }={\bf v}_{1}-\mu _{ji}\left( 1+\alpha
_{ij}^{-1}\right) (\widehat{{\boldsymbol {\sigma }}}\cdot {\bf
g}_{12})\widehat{{\boldsymbol {\sigma }}} ,
\end{equation}
\begin{equation}
 {\bf v}_{2}^{\prime }={\bf v}_{2}+\mu _{ij}\left( 1+\alpha
_{ij}^{-1}\right) (\widehat{{\boldsymbol {\sigma }}}\cdot {\bf
g}_{12})\widehat{ \boldsymbol {\sigma}} ,
\end{equation}
\label{2.3}
\end{subequations}
where $\mu _{ij}\equiv m_{i}/\left( m_{i}+m_{j}\right) $, so that
$\mu_{ij}+\mu_{ji}=1$.}

{However, due to the mathematical complexity of the Boltzmann
equation, and in order to describe general nonequilibrium states, it
is useful to replace the Boltzmann collision operator $J_{ij}\left[
{\bf v}|f_{i},f_{j}\right]$ with a more tractable \emph{model}
operator $K_{ij}\left[ {\bf v}|f_{i},f_{j}\right]$ that reproduces
the collisional transfers of mass, momentum, and energy of the true
inelastic Boltzmann operator, namely
\begin{equation}
\int \dd{\bf v}\,\left\{\begin{array}{l}1\\{\bf
v}\\v^2\end{array}\right\}J_{ij}[{\bf v}|f_{i},f_{j}]=\int \dd{\bf
v}\,\left\{\begin{array}{l}1\\{\bf
v}\\v^2\end{array}\right\}K_{ij}[{\bf v}|f_{i},f_{j}] ,
\label{n2.7}
\end{equation}
 Extending the model proposed by Brey et al.\  \cite{BDS99} for the monocomponent
case and enforcing Eq.\ \eqref{n2.7} in the Gaussian approximation,
 we have recently proposed the following model kinetic equation for
inelastic mixtures \cite{VGS07}:}
\begin{equation}
\partial_t f_i+\mathbf{v}\cdot\nabla f_i=-\sum_{j}\left\{ \frac{1+\alpha_{ij}}{2}\nu_{ij}
\left[f_i(\mathbf{v})-f_{ij}(\mathbf{v})\right]+\frac{\zeta_{ij}}{2}\frac{\partial
}{\partial {\bf v}}\cdot \left[\left( {\bf v
}-\mathbf{u}_{i}\right)f_i(\mathbf{v})\right]\right\},
\label{n5.1.2}
\end{equation}
where
\begin{equation}
\label{5.1}
\nu_{ij}=\frac{4\pi^{(d-1)/2}}{d\Gamma(d/2)}n_j\sigma_{ij}^{d-1}\left(\frac{2\widetilde{T}_i}{m_i}+
\frac{2\widetilde{T}_j}{m_j}\right)^{1/2}
\end{equation}
is a velocity-independent effective collision frequency of a
particle of species $i$ with particles of species $j$,
\begin{equation}
\label{4.1}
\zeta_{ij}=\frac{1}{2}\nu_{ij}\mu_{ji}^2\left[1+\frac{m_i\widetilde{T}_j}{m_j
\widetilde{T}_i}+\frac{3}{2d}\frac{m_i}{\widetilde{T}_i}\left({\bf
u}_i-{\bf u}_j\right)^2\right](1-\alpha_{ij}^2)
\end{equation}
is the contribution to the cooling rate of species $i$ due to the
inelastic collisions with particles of species $j$, and
\begin{equation}
 f_{ij}(\mathbf{v})=n_i\left(\frac{m_i}{2\pi
T_{ij}}\right)^{d/2}
\exp\left[-\frac{m_i}{2T_{ij}}\left(\mathbf{v}-\mathbf{u}_{ij}\right)^2\right]
\label{4.12}
\end{equation}
 is a reference distribution function.
 In the above equations,
\beq
\label{10.1}
\widetilde{T}_i=\frac{m_i}{dn_i}\int \dd{\bf v}\, ({\bf v}-{\bf
u}_i)^2f_i= T_i-\frac{m_i}{d}\left({\bf u}_{i}-{\bf u}\right)^2,
\eeq
\begin{equation}
\mathbf{u}_{ij}=\mu_{ij}\mathbf{u}_{i}+\mu_{ji}\mathbf{u}_{j},
\label{4.16}
\end{equation}
\beq
T_{ij}=\TT_i+2\mu_{ij}\mu_{ji}\left\{\TT_j-\TT_i+\frac{(\mathbf{u}_i-\mathbf{u}_j)^2}{2d}\left[m_j+
\frac{\TT_j-\TT_i}{\TT_i/m_i+\TT_j/m_j}\right]\right\}.
\label{4.18}
\end{equation}

We now specialize to the problem analyzed in this paper, namely a
binary mixture where one of the species ($i=1$) is present in tracer
concentration ($n_1/n_2\to 0$). In this case, Eqs.\ \eqref{2.4} and
\eqref{2.5} imply that $\mathbf{u}=\mathbf{u}_2$ and $T=T_2$. In
addition, the mixture is subjected to the steady Couette flow (see
Fig.\ \ref{fig1}), so that the spatial dependence of all the
quantities is limited to the $y$ variable. In the tracer limit, the
state of the excess component ($i=2$) is not disturbed by the
presence of the impurity and so Eq.\ \eqref{n5.1.2} for $i=2$
becomes
\begin{equation}
\label{1.n}
v_y\frac{\partial f_2}{\partial y}=-\nu_2
(f_2-f_{22})+\frac{\zeta_{22}}{2}\frac{\partial}{\partial {\bf
v}}\cdot \left[\left({\bf v}-\mathbf{u}_2\right) f_2\right],
\end{equation}
where, according to Eqs.\ \eqref{5.1}--\eqref{4.18}, {$\nu_2$,
$\zeta_{22}$, and $f_{22}$ are given by}
\begin{equation}
\label{3.n}
\nu_2=\frac{1+\alpha_{22}}{2}\nu_{22},\quad
\nu_{22}=\frac{8\pi^{(d-1)/2}}{d\Gamma(d/2)}n_2\sigma_{2}^{d-1}\sqrt{\frac{T_2}{m_2}},
\end{equation}
\begin{equation}
\label{4.n}
\zeta_{22}=\frac{1-\alpha_{22}^2}{4}\nu_{22}=\frac{1-\alpha_{22}}{2}\nu_{2},
\end{equation}
\begin{equation}
f_{22}(\mathbf{v})=n_2\left(\frac{m_2}{2\pi T_{2}}\right)^{d/2}
\exp\left[-\frac{m_2(\mathbf{v}-\mathbf{u}_2)^2}{2T_{2}}\right].
\label{2.n}
\end{equation}
Taking moments in Eq.\ \eqref{1.n}, one gets the balance equations
of momentum and energy in the steady state:
\beq
\frac{\partial P_{2,xy}}{\partial y}=\frac{\partial
P_{2,yy}}{\partial y}=0,
\label{II.1}
\eeq
\beq
\frac{\partial q_{2,y}}{\partial y}+\frac{\partial u_{2,x}}{\partial
y}P_{2,xy}=-\frac{d}{2}\zeta_{22}n_2T_2.
\label{II.2}
\eeq

Since the  impurity only collides with particles of the host gas,
Eq.\ \eqref{n5.1.2} for $i=1$ reduces to
\begin{equation}
 v_y\frac{\partial f_1}{\partial y}=-\nu_1
(f_1-f_{12})+\frac{\zeta_{12}}{2}\frac{\partial}{\partial {\bf
v}}\cdot \left[\left({\bf v}-\mathbf{u}_1\right) f_1\right],
\label{16.n}
\end{equation}
where { \beq \nu_1=\frac{1+\alpha_{12}}{2}\nu_{12} \eeq }and
$\nu_{12}$, $\zeta_{12}$, and $f_{12}$ are defined by Eqs.\
\eqref{5.1}--\eqref{4.18} with $\TT_2=T_2=T$. The kinetic equations
\eqref{1.n} and \eqref{16.n} must be supplemented by appropriate
boundary conditions representing the relative motion of the plates
at $y=\pm L/2$.

The main advantage of the tracer limit is that $f_2$ obeys a closed
(inelastic) kinetic equation (the same equation as the monocomponent
granular gas). Once solved, the moments $n_2(y)$, $\mathbf{u}_2(y)$,
and $T_2(y)$ can be inserted into Eq.\ \eqref{16.n} to get a closed
equation for $f_1$. Despite the simplicity of the kinetic model with
respect to the original Boltzmann equation, the search for an exact
solution to the nonlinear Couette flow problem is a formidable task.
In the case of a monocomponent gas, an exact \emph{hydrodynamic}
solution was found in Ref.\ \cite{TTMGSD01}. Of course, this
solution holds for the kinetic equation \eqref{1.n} of the excess
component. Based on this solution, in the next section we obtain an
exact \emph{hydrodynamic} solution for the kinetic equation
\eqref{16.n} of the impurity.

\section{\label{sec3} \protect{Hydrodynamic solution beyond Navier--Stokes order}}
\subsection{Excess component}

As said before, an exact solution to (\ref{1.n}) was found in Ref.\
\cite{TTMGSD01}. Such a solution is characterized by the following
hydrodynamic profiles:
\begin{equation}
\label{III.5} p_2=n_2T_2=\text{const} ,
\end{equation}
\begin{equation}
\label{III.6}
\frac{1}{\nu_2(y)}\frac{\partial}{\partial y}u_{2,x}=a=\text{const},
\end{equation}
\begin{equation}
\label{III.7}
\frac{1}{2m_2}\left[\frac{1}{\nu_2(y)}\frac{\partial}{\partial
y}\right]^2T_{2}=- \gamma(a,\alpha_{22})=\text{const},
\end{equation}
where $\gamma(a,\alpha_{22})\geq 0$ is a dimensionless nonlinear
function of the shear rate $a$ and the coefficient of restitution
$\alpha_{22}$. This quantity  {(henceforth called thermal curvature
coefficient)} characterizes the curvature of the temperature
{profile} as a consequence of both the viscous heating and the
collisional cooling. The form of the profiles
\eqref{III.5}--\eqref{III.7} coincides with the profiles \eqref{5},
\eqref{6}, and \eqref{7} predicted by the NS description, except
that the {thermal curvature coefficient} $\gamma$ differs from {its
NS value} and is determined consistently, {as shown below}. The
solution to Eqs.\ \eqref{III.6} and \eqref{III.7} is
\beq
u_{2,x}(s)=as,\quad T_2(s)=T_2(0)+\epsilon s-m_2\gamma s^2,
\label{III.1}
\eeq
where {the scaled variable $s$ is defined as}
\beq
s(y)=\int_0^y \dd y'\nu_2(y'),
\label{8}
\eeq
{and $\epsilon$
is an arbitrary constant that vanishes if the two wall temperatures
are equal but is nonzero otherwise {($T_{+}\neq T_{-}$)}}
\cite{VU07}.

For convenience, we refer the velocities of the particles to the
Lagrangian frame moving with velocity $u_{2,x}(s)$. In this frame,
Eq.\ \eqref{1.n} can be rewritten as
\beq
\left(1-\frac{d}{2}\zeta_2^*+V_y \partial_s-a
V_y\partial_{V_x}-\frac{1}{2}\zeta_2^*\mathbf{V}\cdot\partial_{\mathbf{V}}\right)f_2(s,\mathbf{V})
=f_{22}(s,\mathbf{V}),
\label{III.2}
\eeq
where
 {
 \beq
 \zeta_2^*=\frac{\zeta_{22}}{\nu_2}=\frac{1-\alpha_{22}}{2}
\eeq
}  and the derivative $\partial_s$ is taken at constant
$\mathbf{V}=\mathbf{v}-\mathbf{u}_2(s)$. Note that the  dependence
of the reference distribution $f_{22}$ on both $s$ and $\mathbf{V}$
is explicit. Taking this into account, the hydrodynamic solution to
Eq.\ \eqref{III.2} is \cite{TTMGSD01}
\beq
f_2(s,\mathbf{V})=\int_0^\infty \dd
w\,e^{-(1-\frac{d}{2}\zeta_2^*)w}e^{-\tau(w,\zeta_2^*)V_y\partial_s}e^{aw
V_y\partial_{V_x}}f_{22}(s,e^{\frac{1}{2}\zeta_2^*w}\mathbf{V}),
\label{III.3}
\eeq
where
 {
 \beq
 \tau(w,\zeta_2^*)\equiv
\frac{2}{\zeta_2^*}\left(e^{\frac{1}{2}\zeta_2^*w}-1\right). \eeq}
The action of the operators $e^{-\tau V_y\partial_s}$ and $e^{aw
V_y\partial_{V_x}}$ on an arbitrary function $g(s,\mathbf{V})$ is
\beq e^{-\tau V_y\partial_s}g(s,\mathbf{V})=g(s-\tau
V_y,\mathbf{V}),\quad e^{aw V_y\partial_{V_x}}g(s,{V}_x)=g(s,V_x+aw
V_y), \label{III.4} \eeq respectively. The solution \eqref{III.3}
clearly adopts the form of a hydrodynamic or \emph{normal} solution
since its spatial dependence only occurs through a functional
dependence on the hydrodynamic fields $n_2(s)$, $\mathbf{u}_2(s)$,
and $T_2(s)$. This provides a neat example of the existence of
normal solutions beyond the NS domain. The solution \eqref{III.3}
depends parametrically on the shear rate $a$, the coefficient of
restitution $\alpha_{22}$ and the thermal curvature {coefficient}
$\gamma$. However, only the two first parameters are independent
since, as indicated by the notation in Eq.\ \eqref{III.7}, $\gamma$
is a nonlinear function of $a$ and $\alpha_{22}$. The parameter
$\gamma(a,\alpha_{22})$ is determined by imposing the consistency
conditions
\begin{equation}
\label{III.8}
\int \dd{\bf v} \{1,{\bf V},V^2\}(f_2-f_{22})=\{0,{\bf 0},0\}.
\end{equation}
While the first two conditions are identically satisfied regardless
of the value of $\gamma$, the  third condition in (\ref{III.8})
leads to the following implicit equation \cite{note1}
\begin{equation}
\label{III.9} d\frac{\zeta_{2}^*}{1+\zeta_{2}^*}-\frac{2a^2}{(1+\zeta_{2}^*)^3}=
2F_{1,0}(\gamma,\zeta_2^*)+dF_{0,0}(\gamma,\zeta_2^*)
+a^2\left[2F_{1,2}(\gamma,\zeta_2^*)+F_{0,2}(\gamma,\zeta_2^*)\right].
\end{equation}
Here, we have introduced the mathematical functions
\begin{equation}
\label{III.12} F_{0,m}(y,z)= \int_{0}^{\infty} \dd w\ e^{-(1+z)w} w^m
\left[\sqrt{\pi}\theta(w,y,z)e^{\theta^2(w,y,z)}\text{erfc}\left(\theta(w,y,z)\right)-
1\right],
\eeq
 {
 \beqa
\label{0.7}
F_{1,m}(y,z)&=& y\frac{\partial}{\partial y} F_{0,m}(y,z)\nn
&=&-\frac{1}{2}\int_{0}^{\infty} \dd w\ e^{-(1+z)w} w^m
\theta(w,y,z)
\left\{\sqrt{\pi}\left[1+2\theta^2(w,y,z)\right]e^{\theta^2(w,y,z)}\text{erfc}\left(\theta(w,y,z)\right)-
2\theta(w,y,z)\right\},\nn
\eeqa}
where $\text{erfc}(x)$ is the complementary error function and
\begin{equation}
\label{III.14}
\theta(w,y,z)=\frac{1}{2\sqrt{2y}} \frac{z}{1- e^{-\frac{1}{2}z w}}.
\end{equation}
The representation (\ref{III.9}) exists only for $\gamma \geq 0$ or,
equivalently, for $a \geq a_\text{th}$, where, as discussed in the
Introduction, the threshold value $a_\text{th}$ of the shear rate
corresponds to $\gamma=0$. In this case,
{$F_{0,m}(0,\zeta_2^*)=F_{1,m}(0,\zeta_2^*)=0$} [see the appendix]
and so
\begin{equation}
\label{III.15} a_\text{th}^2=\frac{d}{2}\zeta_{2}^*(1+\zeta_{2}^*)^2.
\end{equation}
In the case $a=a_\text{th}$ the viscous heating is exactly balanced
by collisional cooling. This state corresponds with the well-known
simple shear flow [if $\epsilon=0$ in \eqref{III.1}] but also to a
non-uniform steady flow  (for $\epsilon\neq 0$)  that has been
reported recently \cite{VU07}.

Once the parameter $\gamma$ is obtained from Eq.\ \eqref{III.9}, the
velocity distribution function is completely determined from Eq.\
\eqref{III.3}. Its relevant moments {provide} the momentum and heat
fluxes. The nonzero elements of the pressure tensor  are given by
\cite{TTMGSD01}
\begin{equation}
\label{III.n1}
\frac{P_{2,xx}}{p_2}=\frac{1}{1+\zeta_{2}^*}+2\frac{a^2}{(1+\zeta_{2}^*)^3}+
F_{0,0}(\gamma,\zeta_2^*)+a^2\left[F_{0,2}(\gamma,\zeta_2^*)+2F_{1,2}(\gamma,\zeta_2^*)\right],
\end{equation}
\begin{equation}
\label{III.n2}
\frac{P_{2,yy}}{p_2}=\frac{1}{1+\zeta_{2}^*}+F_{0,0}(\gamma,\zeta_2^*)+2F_{1,0}(\gamma,\zeta_2^*),
\end{equation}
\begin{equation}
\label{III.n3}
\frac{P_{2,zz}}{p_2}=\frac{1}{1+\zeta_{2}^*}+F_{0,0}(\gamma,\zeta_2^*),
\end{equation}
\begin{equation}
\label{III.n4}
\frac{P_{2,xy}}{p_2}=-a\left[\frac{1}{(1+\zeta_{2}^*)^2}+F_{0,1}(\gamma,\zeta_2^*)+2F_{1,1}(\gamma,\zeta_2^*)\right].
\end{equation}
The requirement $[P_{2,xx}+P_{2,yy}+(d-2)P_{2,zz}]/p_2=d$ is
equivalent to the consistency condition \eqref{III.9}. Equation
\eqref{III.n4} strongly differs from Newton's shearing law {[see
Eq.\ \eqref{4}]} since the quantity enclosed by square brackets in
Eq.\ \eqref{III.n4} is a highly nonlinear function of the shear rate
$a$ through the {thermal curvature coefficient} $\gamma$. For
instance, at $\alpha_{22}=0.8$ and $a=1$ the magnitude of $P_{2,xy}$
is about half its Newtonian value.

Next, we consider the heat flux components $q_{2,x}$ and $q_{2,y}$.
The latter can be easily determined in terms of $P_{2,xy}$ making
use of the energy balance equation \eqref{II.2}, according to which
$q_{2,y}$ is linear in $s$. Consequently, one gets
\beq
q_{2,y}=-\frac{p_2}{2m_2\nu_2\gamma}\left(a\frac{|P_{2,xy}|}{p_2}-\frac{d}{2}\zeta_2^*\right)\frac{\partial
T_2}{\partial y},
\label{III.17}
\eeq
where we have taken into account that $\partial_s T_2$ is also
linear in $s$ [see Eq.\ \eqref{III.1}]. Equation \eqref{III.17} can
be seen as a generalized Fourier's law in the sense that $q_{2,y}$
is proportional to the thermal gradient with an effective thermal
conductivity that is a nonlinear function of the shear rate. The
evaluation of the component $q_{2,x}$ is much more involved.
Multiplying both sides of Eq.\ \eqref{III.3} by $V^2V_x$ and
integrating over velocity, one gets \cite{TTMGSD01}
\beq
q_{2,x}=\frac{p_2}{m_2\nu_2\sqrt{2\gamma}}a\left[G(\gamma,\zeta_2^*)+a^2H(\gamma,\zeta_2^*)\right]\frac{\partial
T_2}{\partial y},
\label{III.18}
\eeq
where we have called
\begin{equation}
\label{III.13}
G(y,z)= \int_{0}^{\infty} \dd w\, e^{-(1+\frac{3}{2}\zeta_{2}^*)w} w
\left[\frac{d+1}{2}X(\theta(w,y,z))+Y(\theta(w,y,z))\right],
\eeq
\begin{equation}
\label{III.13bis}
H(y,z)= \int_{0}^{\infty} \dd w\, e^{-(1+\frac{3}{2}\zeta_{2}^*)w}
w^3 Y(\theta(w,y,z)).
\eeq
Here,
\beq
X(\theta)=\theta^2\left[\sqrt{\pi}(1+2\theta^2)e^{\theta^2}\text{erfc}\left(\theta\right)-2\theta\right],
\label{X}
\eeq
\beq
Y(\theta)=\theta^3\left[2(1+\theta^2)-\sqrt{\pi}\theta(3+2\theta^2)e^{\theta^2}\text{erfc}\left(\theta\right)\right].
\label{Y}
\end{equation}
The existence of  a component of the heat flux orthogonal to the
thermal gradient and parallel to the flow direction goes beyond
Fourier's law. In fact, $q_{2,x}$ is at least of order $a (\partial
T_2/\partial y)$ and so Eq.\ \eqref{III.18} can be seen as a
generalized Burnett effect.

\subsection{Impurity particle}
Once the hydrodynamic state of the excess component has been
characterized, we next want to analyze the hydrodynamic state of the
impurity particle.

First, some useful information can be extracted by taking moments in
Eq.\ \eqref{16.n}: \beq \frac{\partial P_{1,yy}}{\partial y}=0,
\label{III.19} \eeq \beq \frac{\partial P_{1,xy}}{\partial
y}=-\nu_1\rho_1\left(u_{1,x}-u_{12,x}\right), \label{III.20} \eeq
\beq \frac{\partial q_{1,y}}{\partial y}+\frac{\partial
u_{2,x}}{\partial
y}P_{1,xy}=-\nu_1\left[\frac{d}{2}n_1(T_1-T_{12})-\frac{\rho_1}{2}(\mathbf{u}_{12}-\mathbf{u}_2)^2\right]
-\frac{d}{2}\zeta_{12}n_1\left[T_1-\frac{m_1}{d}(\mathbf{u}_1-\mathbf{u}_2)^2\right].
\label{III.21} \eeq Next, we guess (to be confirmed later) that the
hydrodynamic state of the impurity is enslaved to that of the
granular gas in the sense that {
\begin{description}
\item[](i) there is no mutual diffusion, i.e., ${\bf u}_1(y)={\bf u}_2(y)$,
\item[](ii) the mole fraction $n_1(y)/n_2(y)$ is uniform, and
\item[](iii) the temperature ratio $\chi\equiv T_1(y)/T_2(y)$ is also
uniform.
\end{description}
} The latter parameter $\chi$ must be a function of the {mass and
size ratios
 \beq
 \mu=\frac{m_1}{m_2},\quad \omega=\frac{\sigma_1}{\sigma_2},
 \eeq}
 the reduced shear rate $a$, and the
coefficients of restitution $\alpha_{22}$ and $\alpha_{12}$. Of
course, the temperature ratio is $\chi=1$ when the impurity is
mechanically equivalent to the gas particles ($\mu=\omega=1$,
$\alpha_{12}=\alpha_{22}$). Taking into account  assumption (i),
Eqs.\ \eqref{III.20} and \eqref{III.21} become \beq \frac{\partial
P_{1,xy}}{\partial y}=0, \label{III.22} \eeq \beq {\partial_s
q_{1,y}}+aP_{1,xy}=-\frac{d}{2}n_1 T_1\frac{\nu_1}{\nu_2}
\left(1-\frac{T_{12}}{T_1}+\frac{\zeta_{12}}{\nu_1}\right).
\label{III.23} \eeq {Furthermore,  assumptions (ii) and (iii) imply
that} the product $n_1T_1$ and the ratios $T_{12}/T_1$,
$\nu_1/\nu_2$, and $\zeta_{12}/\nu_1$ are constant quantities. The
three latter are given by \beq
\frac{T_{12}}{T_1}=1+\frac{2\mu(1-\chi)}{(1+\mu)^2\chi},
\label{III.24a} \eeq
\begin{equation}
\label{III.18n}
\frac{\nu_{1}}{\nu_2}=\frac{1+\alpha_{12}}{1+\alpha_{22}}
\left(\frac{1+\omega}{2}\right)^{d-1}\sqrt{\frac{\mu+\chi}{2\mu}},
\end{equation}
\begin{equation}
\label{III.19n}
\zetat\equiv\frac{\zeta_{12}}{\nu_1}=
\frac{\mu+\chi}{(1+\mu)^2\chi}(1-\alpha_{12}).
\end{equation}

{}From a formal point of view, the kinetic equation (\ref{1.n})
becomes Eq.\ (\ref{16.n}) by making the changes $f_2 \to f_1$,
$f_{22}\to f_{12}$, $\nu_2\to \nu_1$ and $\zeta_{22}\to \zeta_{12}$.
The formal change  $f_{22}\to f_{12}$ implies the changes $n_2\to
n_1$, $m_2\to m_1$, and $T_2\to T_{12}$. It is then convenient to
introduce the {auxiliary} quantities
\begin{equation}
\label{III.21n}
\widetilde{a}=\frac{1}{\nu_1(y)}\frac{\partial}{\partial
y}u_{2,x}=a\frac{\nu_2}{\nu_1},
\end{equation}
\begin{equation}
\label{III.22n}
\widetilde{\gamma}=-\frac{1}{2m_1}\left[\frac{1}{\nu_1(y)}\frac{\partial}{\partial
y}\right]^2T_{12}=\left(\frac{\nu_2}{\nu_1}\right)^2\frac{T_{12}}{T_1}\frac{\chi}{\mu}\gamma.
\end{equation}
Equations (\ref{III.21n}) and (\ref{III.22n}), along with $n_1
T_{12}=\text{const}$, define the profiles of the fields
characterizing the distribution function $f_{12}$.

The formal mapping described above allows us to easily get the
moments of $f_1$  from comparison with those of $f_2$. In
particular, the two first self-consistency conditions are verified,
namely
\begin{equation}
\label{III.8bis}
\int \dd{\bf v} \{1,{\bf V}\}(f_1-f_{12})=\{0,{\bf 0}\},
\end{equation}
regardless of the values of $\gamma$ and $\chi$. The third
self-consistency condition reads
\beq
\frac{m_1}{d}\int \dd{\bf v} V^2(f_1-f_{12})=n_1
T_1\left(1-\frac{T_{12}}{T_1}\right). \label{III.23b}
\end{equation}
This condition determines the temperature ratio $\chi$. To evaluate
the left-hand side of Eq.\ \eqref{III.23b}, it is convenient to
obtain first the nonzero elements of the partial pressure tensor
$\mathsf{P}_1$. {}Taking into account  Eqs.\
\eqref{III.n1}--\eqref{III.n4}, one gets
\begin{equation}
\label{n5}
\frac{P_{1,xx}}{n_1T_{12}}=\frac{1}{1+\zetat}+2\frac{\widetilde{a}^2}{(1+\zetat)^3}+
F_{0,0}(\widetilde{\gamma},\zetat)+\widetilde{a}^2\left[F_{0,2}(\widetilde{\gamma},\zetat)
+2F_{1,2}(\widetilde{\gamma},\zetat)\right]
,
\end{equation}
\begin{equation}
\label{n6}
\frac{P_{1,yy}}{n_1T_{12}}=\frac{1}{1+\zetat}+F_{0,0}(\widetilde{\gamma},\zetat)+2F_{1,0}(\widetilde{\gamma},\zetat)
,
\end{equation}
\begin{equation}
\label{n7} \frac{P_{1,zz}}{n_1T_{12}}=\frac{1}{1+\zetat}+F_{0,0}(\widetilde{\gamma},\zetat) ,
\end{equation}
\begin{equation}
\label{n8}
\frac{P_{1,xy}}{n_1T_{12}}=-\frac{\widetilde{a}}{(1+\zetat)^2}-\widetilde{a}F_{0,1}(\widetilde{\gamma},\zetat)-
2\widetilde{a}F_{1,1}(\widetilde{\gamma},\zetat),
\end{equation}
where the functions {$F_{0,m}(y,z)$ and $F_{1,m}(y,z)$ are defined
by Eqs.\ \eqref{III.12} and \eqref{0.7}, respectively}. Condition
\eqref{III.23b} is equivalent to
$P_{1,xx}+P_{1,yy}+(d-2)P_{1,zz}=dn_1T_1$, yielding
\begin{equation}
\label{n9}
d\left(\frac{T_1}{T_{12}}-\frac{1}{1+\zetat}\right)-\frac{2\widetilde{a}^2}
{(1+\zetat)^3}=2F_{1,0}(\widetilde{\gamma},\zetat)+dF_{0,0}(\widetilde{\gamma},\zetat)
+\widetilde{a}^2\left[2F_{1,2}(\widetilde{\gamma},\zetat)+F_{0,2}(\widetilde{\gamma},\zetat)\right].
\end{equation}
For given values of the reduced shear rate $a$ and the mechanical
parameters of the system ($\alpha_{22}$, $\alpha_{12}$, $\mu$, and
$\omega$),  Eq.\ \eqref{n9}, complemented with Eq.\ \eqref{III.9}
and the relations (\ref{III.24a})--(\ref{III.22n}), becomes a
nonlinear closed equation for  the temperature ratio $\chi$, which
must be {solved} numerically. In the case of mechanically equivalent
particles, Eq.\ \eqref{n9} yields $\chi=1$ and is equivalent to Eq.\
\eqref{III.9}. Insertion of this solution into Eqs.\
\eqref{n5}--\eqref{n8} gives the elements of the pressure tensor
$\mathsf{P}_1$.

We consider now the heat flux associated with the impurity.
According to Eq.\ \eqref{III.23}, $q_{1,y}$ is linear in $s$. Since
$\partial_s T_2$ is also linear in $s$ [cf.\ Eq.\ \eqref{III.1}],
one can write
\beq
q_{1,y}=-\frac{n_1T_1}{2m_2\nu_2\gamma}\left[a\frac{|P_{1,xy}|}{n_1T_1}-\frac{d}{2}\frac{\nu_1}{\nu_2}
\left(1-\frac{T_{12}}{T_1}+\zetat\right)\right]\frac{\partial
T_2}{\partial y}.
\label{III.24}
\eeq
To get the $x$-component of the heat flux, we make use again of the
formal mapping described above. Thus, from Eq.\ \eqref{III.18} we
obtain
\beq
q_{1,x}=\frac{n_1T_{12}}{m_1\nu_1\sqrt{2\widetilde{\gamma}}}\widetilde{a}
\left[G(\widetilde{\gamma},\zetat)+\widetilde{a}^2
H(\widetilde{\gamma},\zetat)\right]\frac{\partial T_{12}}{\partial
y}.
\label{III.25}
\eeq

\subsection{Generalized transport coefficients for the impurity
particle} In order to characterize the momentum and heat transport
associated with the impurity particle we introduce {five}
generalized transport coefficients. The shear stress $P_{1,xy}$
defines a (dimensionless) nonlinear shear viscosity coefficient
$\eta_1$ as \beq P_{1,xy}=-\eta_1\frac{n_1 T_2}{\nu_1}\frac{\partial
u_{2,x}}{\partial y}. \label{III.26} \eeq The anisotropy of the
normal stresses can be measured through the {viscometric}
coefficients $N_1$ and $M_1$: \beq
\frac{P_{1,xx}-P_{1,yy}}{n_1T_1}=N_1, \quad
\frac{P_{1,zz}-P_{1,yy}}{n_1T_1}=M_1. \label{III.27} \eeq The heat
flux defines a generalized thermal conductivity coefficient
$\lambda_1$ and a cross coefficient $\phi_1$ as \beq
q_{1,y}=-\lambda_1\frac{d+2}{2}\frac{n_1T_2}{m_1\nu_1}
\frac{\partial T_2}{\partial y}, \quad
q_{1,x}=\phi_1\frac{d+2}{2}\frac{n_1T_2}{m_1\nu_1} \frac{\partial
T_2}{\partial y}. \label{III.28} \eeq {}From Eqs.\
\eqref{n5}--\eqref{n8}, \eqref{III.24}, and \eqref{III.25} it is
possible to identify the expressions for these five generalized
transport coefficients. They are given by
\begin{equation}
\label{III.29}
\eta_1=\frac{T_{12}}{T_1}\chi\left[\frac{1}{(1+\zetat)^2}+F_{0,1}(\widetilde{\gamma},\zetat)+
2F_{1,1}(\widetilde{\gamma},\zetat)\right],
\end{equation}
\begin{equation}
\label{III.30}
N_1=\frac{T_{12}}{T_1}\left\{2\frac{\widetilde{a}^2}{(1+\zetat)^3}
+\widetilde{a}^2\left[F_{0,2}(\widetilde{\gamma},\zetat)+
2F_{1,2}(\widetilde{\gamma},\zetat)\right]-2F_{1,0}(\widetilde{\gamma},\zetat)\right\}
,
\end{equation}
\begin{equation}
\label{III.31}
M_1=-2\frac{T_{12}}{T_1}F_{1,0}(\widetilde{\gamma},\zetat) ,
\end{equation}
\beq
\lambda_1=\frac{1}{d+2}\frac{T_{12}}{T_1}\frac{\chi^2}{\widetilde{\gamma}}\left[\eta_1\widetilde{a}^2-\frac{d}{2}
\left(1-\frac{T_{12}}{T_1}+\zetat\right)\right], \label{III.32} \eeq
\beq
\phi_1=\frac{2}{d+2}\left(\frac{T_{12}}{T_1}\right)^2\frac{\chi^2}{\sqrt{2\widetilde{\gamma}}}
\widetilde{a}\left[G(\widetilde{\gamma},\zetat)+\widetilde{a}^2
H(\widetilde{\gamma},\zetat)\right]. \label{III.33} \eeq {Their
expressions}  in the limit $a\to a_{\text{th}}$ are explicitly given
in the appendix.

\section{Monte Carlo simulations\label{sec4}}
As said before, the exact solution to the kinetic equation
\eqref{16.n} derived in Sec.\ \ref{sec3} defines a normal or
hydrodynamic solution where its spatial dependence only occurs
through the hydrodynamic fields ($n_1$, $n_2$, $\mathbf{u}_2$, and
$T_2$) and their gradients. This solution is free from
boundary-layer effects and formally corresponds to idealized
boundary conditions of infinitely cold walls ($T_w\to 0$). For more
details the reader is referred to Appendix B of Ref.\
\cite{TTMGSD01}. The important point is whether or not this exact
solution actually describes the steady state reached by the system,
in the bulk domain, when subject to realistic boundary conditions
and for arbitrary initial conditions. To confirm this expectation,
one needs to solve numerically the set of time-dependent kinetic
equations
\begin{equation}
\partial_t f_i+ v_y\frac{\partial f_i}{\partial y}=-\nu_i
(f_i-f_{i2})+\frac{\zeta_{i2}}{2}\frac{\partial}{\partial {\bf
v}}\cdot \left({\bf v}-\mathbf{u}_i\right) f_i\,\quad i=1,2.
\label{MC1}
\end{equation}
These equations are solved with boundary conditions at $y=\pm
L/2$ compatible with the wall values $\pm U/2$ and $T_w$ and
starting from an arbitrary initial condition. Specifically, we have
considered Maxwellian diffuse boundary conditions
\cite{TTMGSD01,MSG00} and an initial distribution of total
equilibrium. The latter choice does not imply a loss of generality
in the base steady states that are achieved in the system and only
affects  the transient evolution.  Both species are let to
simultaneously evolve from the initial state.  It is also to be
noticed that in the numerical solution of Eq.\ \eqref{MC1} there is
no \textit{a priori} assumption of equal flow velocities for the two
components. i.e., eventual steady-state solutions with
$\mathbf{u}_1\neq\mathbf{u}_2$ are let to occur. However, as we will
show, this actually never happens and all the steady states found
are consistent with $\mathbf{u}_1=\mathbf{u}_2$ (absence of
diffusion).

In this paper we have employed a direct simulation Monte Carlo
(DSMC) method \cite{Bird,AG97} to numerically solve the kinetic
equations \eqref{MC1} in the three-dimensional case. The DSMC method
has been extensively used to solve kinetic equations like the
Boltzmann  and BGK  equations and it has {proven} to accurately
describe transport phenomena in elastic gases and has also
successfully been extended to flows in granular gases. In the DSMC
method two steps are taken every time interval $\delta t$: the free
streaming step, during which a particle with velocity $\mathbf{v}$
is drifted by $\mathbf{v}\delta t$ and the boundary conditions are
applied to those particles leaving the system, and the collision
step, in which $\nu_{i}\delta t$ collision pairs are randomly
selected among neighbor particles, $\nu_i$ being the characteristic
collision frequency in the kinetic equation. Our method  differs
from the elastic case in the addition, in the free streaming step,
of the drag term which mimics the inelasticity in the collisions.

The distributions $f_i$ are represented by $\mathcal{N}_i$ particles
with velocities $\{\mathbf{v}_k\}$ and positions $\{y_k\}$,
$k=1,\ldots,\mathcal{N}_i$. The system is split into $\mathcal{M}$
layers $I=1,\ldots \mathcal{M}$ of width $\delta y=L/\mathcal{M}$.
The particles with positions belonging in layer $I$ define the
densities $n_{i,I}$, the flow velocities $\mathbf{u}_{i,I}$, and the
temperatures $T_{i,I}$ of that layer. {}From those quantities one
can evaluate $\nu_{i,I}$ and $\zeta_{i2,I}$. The free streaming and
the collision steps are briefly described below.

\subsection{Free streaming}
In the free streaming step the positions and velocities for  both
components are updated with the following rules:
\begin{eqnarray}
y_k &\to & y_k+v_{k,y}\delta t, \nonumber
\\
\mathbf{v}_k & \to & \mathbf{u}_{i,I}+e^{-\zeta_{i2,I}\delta
t/2}\left(\mathbf{v}_k-\mathbf{u}_{i,I}\right),
\label{MC2}\end{eqnarray} where $I$ is the layer the particle $k$
belongs {in}. The spatial and velocity updates \eqref{MC2}
 are valid as long as the particle does not leave the
system, i.e., $|y_k+v_{k,y}\delta t|<L/2$. Otherwise, the particle
is reentered by applying thermal boundary conditions. If the
particle ``crosses'' a wall, then
\beq
\mathbf{v}_k\to \pm(U/2)\widehat{\mathbf{x}}+\mathbf{w}_k,
\eeq
where the velocity components $w_{k,x}, w_{k,z}$ are randomly picked
from Maxwell distribution functions (at a temperature $T_w$) whereas
$w_{k,y}=\mp\upsilon$ (upper and lower signs for top and bottom wall
collision, respectively) with $\upsilon>0$ being a random velocity
sampled from the Rayleigh probability distribution
\beq
P_{i}(\mathbf{\upsilon})=\frac{m_i \upsilon} {T_w}e^{-
m_i\upsilon^2/2T_w}.
\eeq
The new position after wall collision is
\beq
y_i\to \pm L/2+w_{k,y}\left(\delta t-\frac{\pm
L/2-y_k}{v_{k,y}}\right).
\eeq

\subsection{Collision step}
For each layer $I$  a number $\nu_{i,I}\delta t$ of particles is
randomly selected among those belonging in the layer. Then the
velocity $\mathbf{v}_k$ of each one of those particles is replaced
by
\beq
\mathbf{v}_k\to \mathbf{u}_{i,I}+\mathbf{V}_k,
\eeq
where $\mathbf{V}_k$ is a random velocity sampled from a Maxwell
probability distribution, with temperatures $T_{12}$ and $T_2$
  for species $i=1$ and $i=2$, respectively.

\subsection{Time and length scales and simulation technical facts}
In the simulations, the quantities are reduced  using
$\overline{\ell}$ and $\overline{\tau}$ as  length and time units,
respectively, where
\beq
\overline{\ell}=
\frac{3}{4(1+\alpha_{22})}\frac{1}{\sqrt{2\pi}\overline{n}_2\sigma_2^2},
\quad \overline{\tau}=\frac{\overline{\ell}}{v_0},
\eeq
 $\overline{n}_2$ and $v_0=\sqrt{2T_w/m_2}$ being the average density of the gas
 particles and a reference thermal velocity, respectively.

Since the aim of DSMC simulations is to solve a kinetic equation, it
must be able to describe the dynamical processes occurring in the
system at a microscopic level \cite{Bird}. This means that the width
layer $\delta y$  must be small compared to the typical microscopic
length scale, determined by the mean free path $\ell_i$. Similarly,
the time step $\delta t$  must be small compared to the inverse of
the collision frequency, $\nu_i^{-1}$. Also, for obtaining an
ergodic simulation, the number of simulated particles $N_i$ must be
sufficiently large. We therefore have performed simulations, for
both species, with $N_i=2\times 10^6$ particles, $\delta y=2\times
10^{-2}\overline\ell$, and $\delta t=3\times
10^{-3}\overline{\tau}$. In order to probe a nonlinear Couette flow
with $\gamma>0$ ($a>a_{\text{th}}$), we have taken a wall velocity
difference $U=10 v_0$ and a system size typically in the range
$L\approx 2$--$20 \overline{\ell}$, {which produces sufficiently
large values of $a$}.

{Taking into account that} the relation between microscopic over
hydrodynamic scales is given by the Knudsen number \textrm{Kn}, the
bin $\delta y_h$ we pick for measurements of the hydrodynamic
profiles, including transport coefficients, is of the order of
$\delta y_h= 0.2\textrm{Kn}^{-1} {\ell_i}$ (note that, in our
system, the reduced local shear rate $a$ is the reference measure
for the Knudsen number). This means that the  measurements of the
hydrodynamic properties are performed over sets of microscopic
cells, i.e., an average over microscopic cells is taken for each set
(macroscopic cell). In this way, the fluctuations of  macroscopic
magnitudes, typical in DSMC simulations, are greatly reduced and
profiles are smoothed with no loss of resolution at a hydrodynamic
scale. Prior to averaging over sets of cells, the hydrodynamic
quantities and fluxes  are obtained for each cell, by using the
expressions that may be found in Ref.\ \cite{MSG00}.

As already explained in Secs.\ \ref{sec1} and \ref{sec3}, the
reduced shear rate $a$ and the thermal curvature {coefficient}
$\gamma$ are fundamental quantities in the problem. We measured
these quantities by fitting the velocity and temperature profiles
from the simulations to fourth-degree polynomials and extracting
from these fits the derivatives appearing in the expressions
\eqref{III.6} and \eqref{III.7}.

An important point in DSMC simulations is the quality of the random
number generator. We used for this purpose random generators from
Intel MKL 9.1 \cite{IMKL}, whose performance has been rigorously
examined in technical tests. The DSMC code was written in C language
and compiled with Intel C++ 10.0 compiler and run in 64-bit Linux
machines.

\section{Results\label{sec5}}
\begin{figure}
\includegraphics[width=.5\columnwidth]{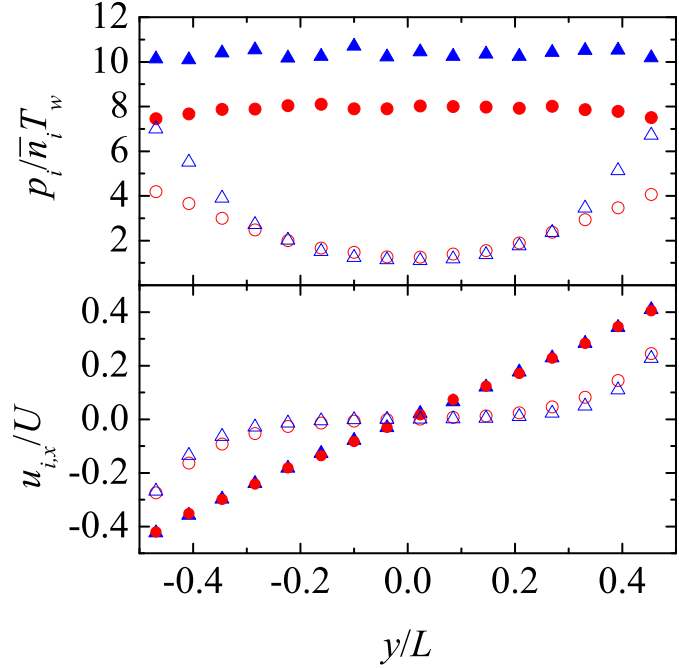}
\caption{(Color online) Pressure and flow velocity profiles for the
impurity (triangles) and the gas particles (circles) at $t\approx
\overline{\tau}$ (open symbols) and $t\approx 10^3 \overline{\tau}$
(filled symbols). The system corresponds to
$\alpha_{12}=\alpha_{22}=0.9$, $m_1/m_2=2$, $\sigma_{1}/\sigma_2=1$,
$L=3.25\overline{\ell}$, and $U=10 v_0$. {At short times}, the
hydrostatic pressures $p_i$ are not constant and the flow velocities
$\mathbf{u}_i$ are not equal, but the simulation quickly evolves to
$p_1= \text{const}$, $p_2=\text{const}$, and
$\mathbf{u}_1=\mathbf{u}_2$, just like in the theoretical solution.
We observed analogous evolutions in all simulations we performed,
for a wide range of parameter values.}
\label{profiles}
\end{figure}

\begin{figure}
\includegraphics[width=.5\columnwidth]{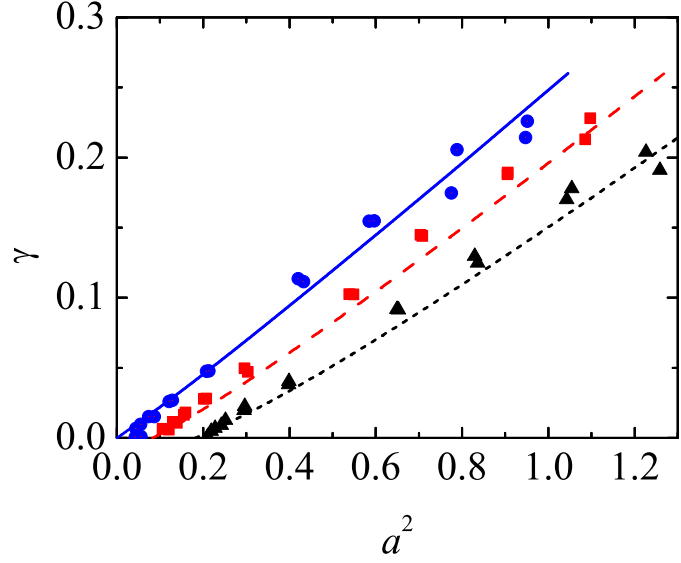}
\caption{(Color online) Shear rate dependence of the parameter
$\gamma$ measuring the curvature of the temperature profile [see
Eq.\ \protect\eqref{III.7}] for $\alpha_{22}=1$ (solid line and
circles), $\alpha_{22}=0.9$ (dashed line and squares), and
$\alpha_{22}=0.8$ (dotted line and triangles). The symbols represent
the simulation results, while the lines are the theoretical
predictions given by Eq.\ \protect\eqref{III.9}.}
\label{gamma}
\end{figure}

\begin{figure}
\includegraphics[width=.5\columnwidth]{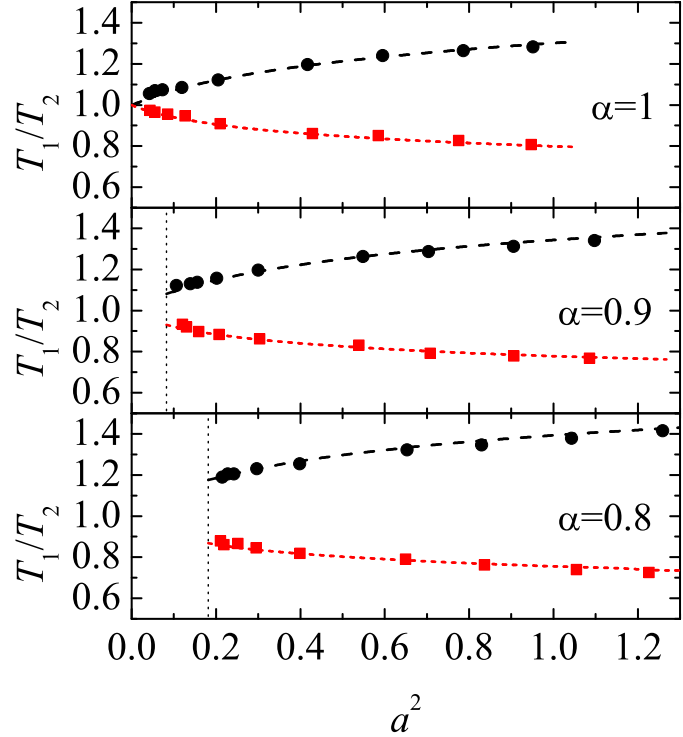}
\caption{(Color online) Shear rate dependence of the temperature
ratio $\chi\equiv T_1/T_2$ in the case of an impurity particle with
$\omega\equiv\sigma_{1}/\sigma_{2}=1$,
$\alpha_{11}=\alpha_{22}=\alpha$, and $\mu\equiv m_1/m_2=2$ (dashed
lines and circles) and $\mu\equiv m_1/m_2=1/2$ (dotted lines and
squares). The symbols represent the simulation results, while the
lines are the theoretical predictions given by Eq.\
\protect\eqref{n9}. The top, middle, and bottom panels correspond to
$\alpha=1$, $\alpha=0.9$, and $\alpha=0.8$, respectively. The dotted
vertical lines indicate the location of the threshold value
$a_{\text{th}}^2(\alpha)$.}
\label{chi}
\end{figure}

\begin{figure}
\includegraphics[width=.5\columnwidth]{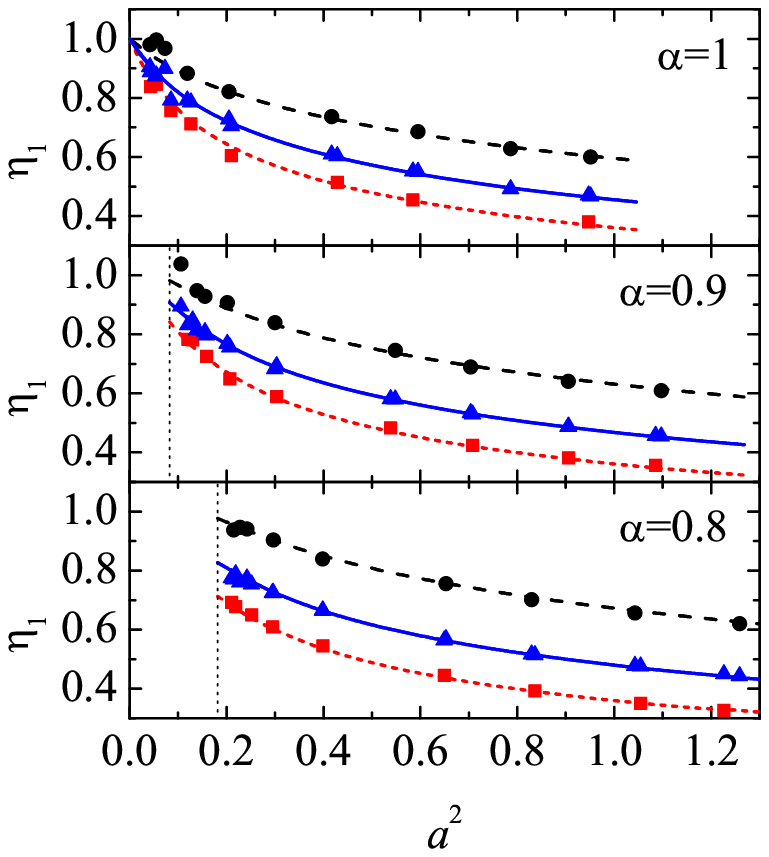}
\caption{(Color online) Shear rate dependence of the nonlinear shear
viscosity coefficient $\eta_1$ [see Eq.\ \protect\eqref{III.26}]
associated with an impurity particle with
$\omega\equiv\sigma_{1}/\sigma_{2}=1$,
$\alpha_{11}=\alpha_{22}=\alpha$, and $\mu\equiv m_1/m_2=2$ (dashed
lines and circles), $\mu\equiv m_1/m_2=1$ (solid lines and
triangles), and $\mu\equiv m_1/m_2=1/2$ (dotted lines and squares).
The symbols represent the simulation results, while the lines are
the theoretical predictions given by Eq.\ \protect\eqref{III.29}.
The top, middle, and bottom panels correspond to $\alpha=1$,
$\alpha=0.9$, and $\alpha=0.8$, respectively. The dotted vertical
lines indicate the location of the threshold value
$a_{\text{th}}^2(\alpha)$.}
\label{eta1}
\end{figure}

\begin{figure}
\includegraphics[width=.5\columnwidth]{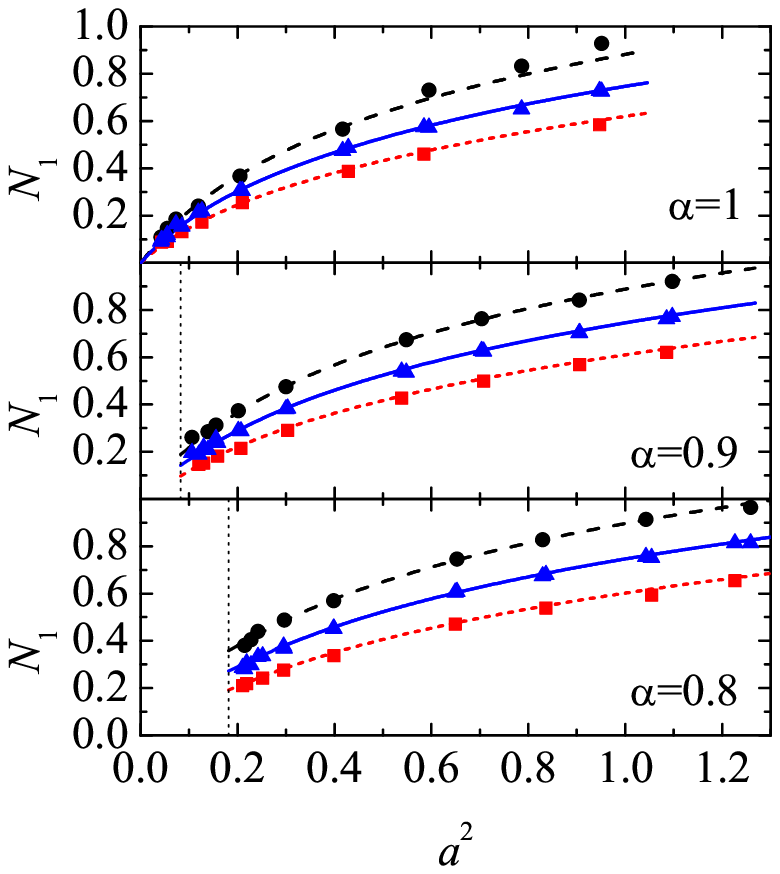}
\caption{(Color online) Shear rate dependence of the reduced normal
stress difference $N_1$ [see Eq.\ \protect\eqref{III.27}] associated
with an impurity particle with
$\omega\equiv\sigma_{1}/\sigma_{2}=1$,
$\alpha_{11}=\alpha_{22}=\alpha$, and $\mu\equiv m_1/m_2=2$ (dashed
lines and circles), $\mu\equiv m_1/m_2=1$ (solid lines and
triangles), and $\mu\equiv m_1/m_2=1/2$ (dotted lines and squares).
The symbols represent the simulation results, while the lines are
the theoretical predictions given by Eq.\ \protect\eqref{III.30}.
The top, middle, and bottom panels correspond to $\alpha=1$,
$\alpha=0.9$, and $\alpha=0.8$, respectively. The dotted vertical
lines indicate the location of the threshold value
$a_{\text{th}}^2(\alpha)$.}
\label{N1}
\end{figure}

\begin{figure}
\includegraphics[width=.5\columnwidth]{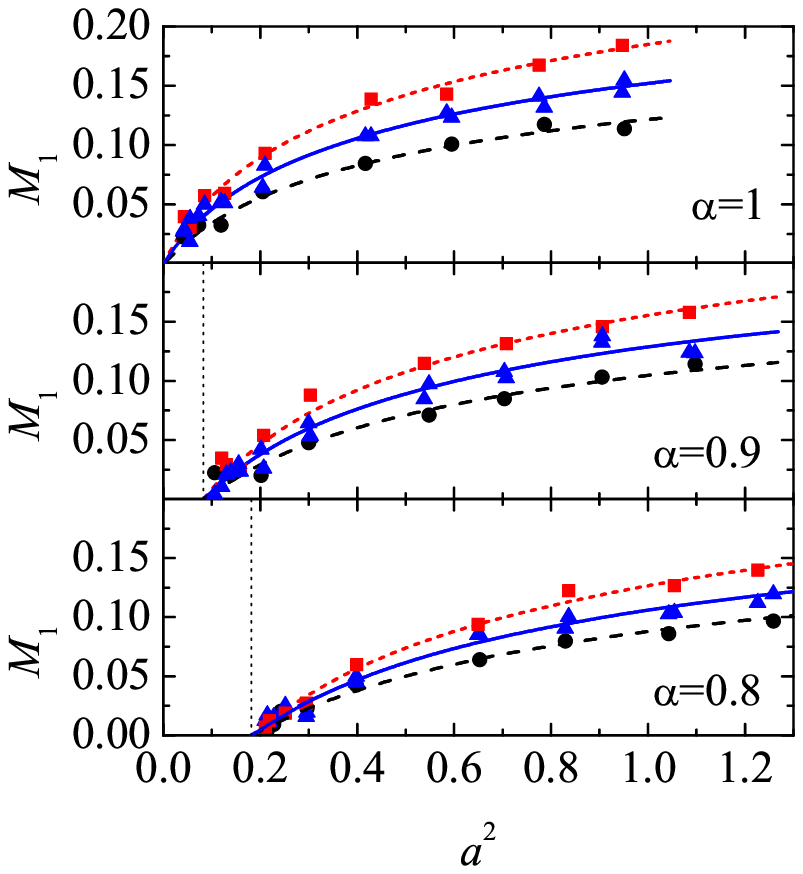}
\caption{(Color online) Shear rate dependence of the reduced normal
stress difference $M_1$ [see Eq.\ \protect\eqref{III.27}] associated
with an impurity particle with
$\omega\equiv\sigma_{1}/\sigma_{2}=1$,
$\alpha_{11}=\alpha_{22}=\alpha$, and $\mu\equiv m_1/m_2=2$ (dashed
lines and circles), $\mu\equiv m_1/m_2=1$ (solid lines and
triangles), and $\mu\equiv m_1/m_2=1/2$ (dotted lines and squares).
The symbols represent the simulation results, while the lines are
the theoretical predictions given by Eq.\ \protect\eqref{III.31}.
The top, middle, and bottom panels correspond to $\alpha=1$,
$\alpha=0.9$, and $\alpha=0.8$, respectively. The dotted vertical
lines indicate the location of the threshold value
$a_{\text{th}}^2(\alpha)$.}
\label{M1}
\end{figure}

\begin{figure}
\includegraphics[width=.5\columnwidth]{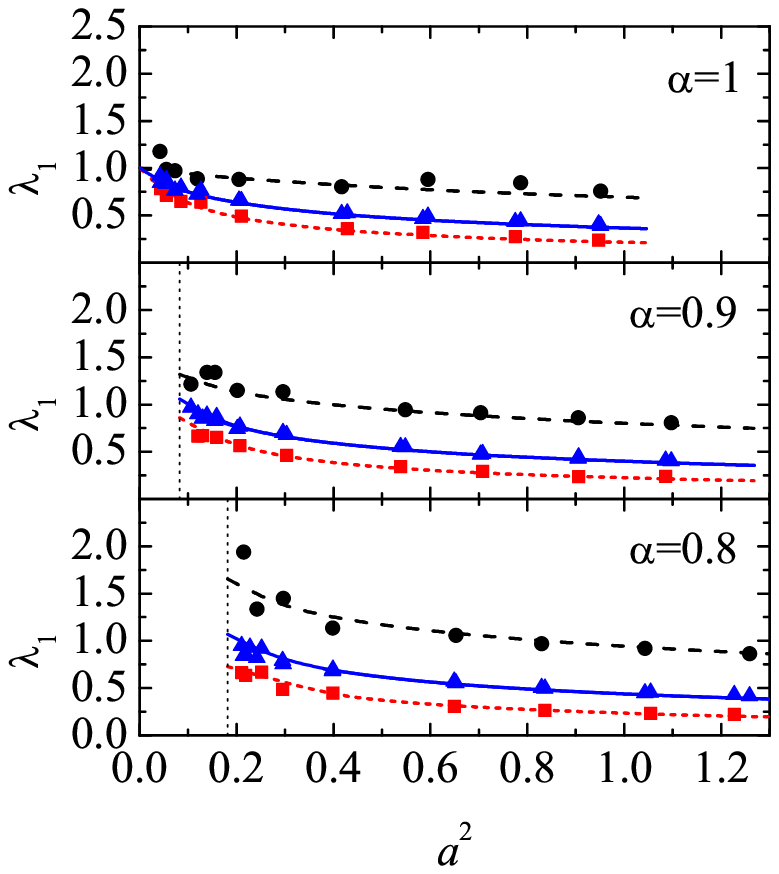}
\caption{(Color online) Shear rate dependence of the nonlinear
thermal conductivity coefficient $\lambda_1$ [see Eq.\
\protect\eqref{III.28}] associated with an impurity particle with
$\omega\equiv\sigma_{1}/\sigma_{2}=1$,
$\alpha_{11}=\alpha_{22}=\alpha$, and $\mu\equiv m_1/m_2=2$ (dashed
lines and circles), $\mu\equiv m_1/m_2=1$ (solid lines and
triangles), and $\mu\equiv m_1/m_2=1/2$ (dotted lines and squares).
The symbols represent the simulation results, while the lines are
the theoretical predictions given by Eq.\ \protect\eqref{III.32}.
The top, middle, and bottom panels correspond to $\alpha=1$,
$\alpha=0.9$, and $\alpha=0.8$, respectively. The dotted vertical
lines indicate the location of the threshold value
$a_{\text{th}}^2(\alpha)$.}
\label{lambda1}
\end{figure}

\begin{figure}
\includegraphics[width=.5\columnwidth]{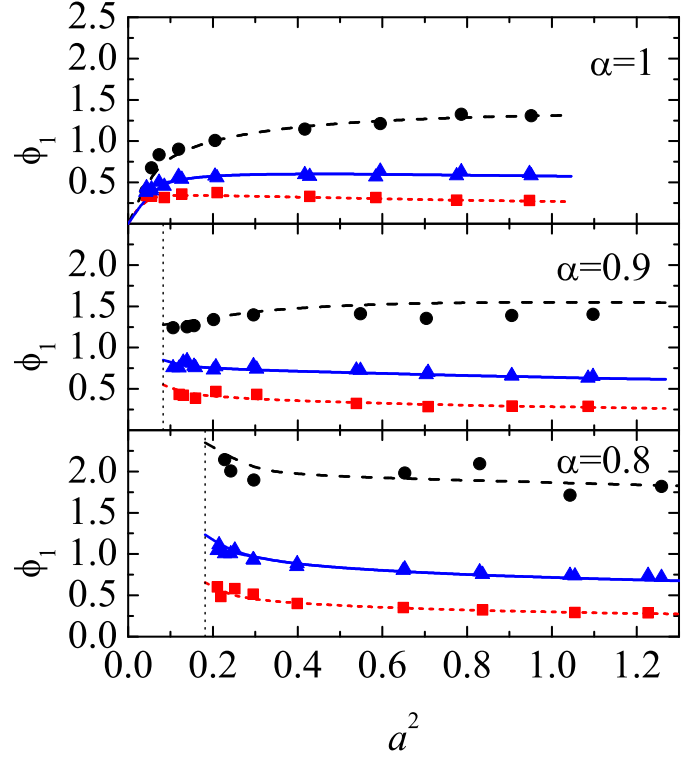}
\caption{(Color online) Shear rate dependence of the cross
coefficient $\phi_1$ [see Eq.\ \protect\eqref{III.28}] associated
with an impurity particle with
$\omega\equiv\sigma_{1}/\sigma_{2}=1$,
$\alpha_{11}=\alpha_{22}=\alpha$, and $\mu\equiv m_1/m_2=2$ (dashed
lines and circles), $\mu\equiv m_1/m_2=1$ (solid lines and
triangles), and $\mu\equiv m_1/m_2=1/2$ (dotted lines and squares).
The symbols represent the simulation results, while the lines are
the theoretical predictions given by Eq.\ \protect\eqref{III.33}.
The top, middle, and bottom panels correspond to $\alpha=1$,
$\alpha=0.9$, and $\alpha=0.8$, respectively. The dotted vertical
lines indicate the location of the threshold value
$a_{\text{th}}^2(\alpha)$.}
\label{phi1}
\end{figure}

\begin{figure}
\includegraphics[width=.5\columnwidth]{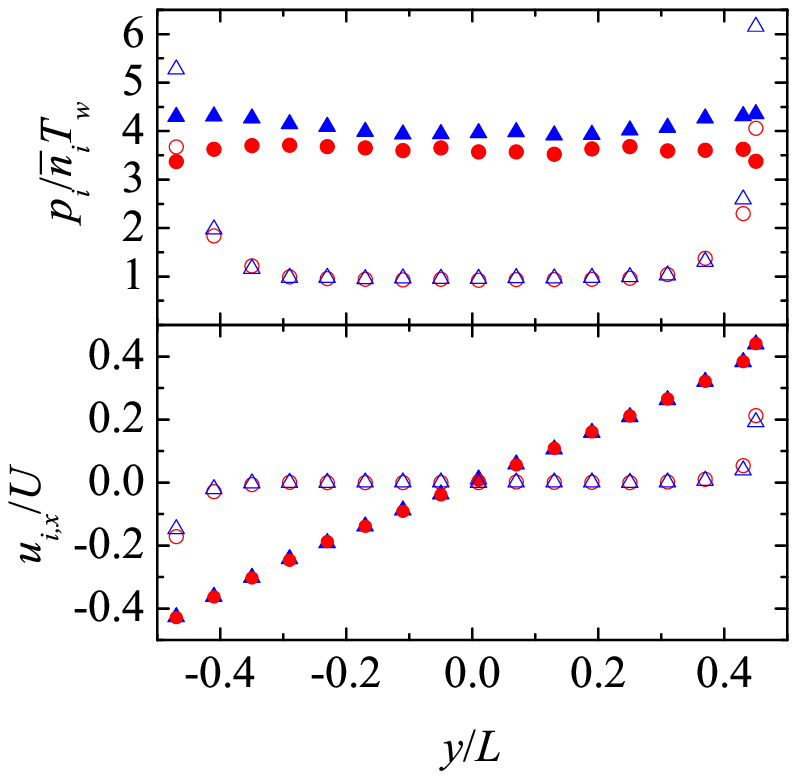}
\caption{{(Color online) Pressure and flow velocity profiles
for the impurity (triangles) and the gas particles (circles) at
$t\approx \overline{\tau}$ (open symbols) and $t\approx 10^3
\overline{\tau}$ (filled symbols). The system corresponds to
$\alpha_{12}=\alpha_{22}=0.9$, $m_1/m_2=2$, $\sigma_{1}/\sigma_2=1$,
$L=7.14\overline{\ell}$, and $U=10 v_0$. The data have been obtained
from DSMC simulations of the Boltzmann equation.}}
\label{profiles2}
\end{figure}

\subsection{{Enslaving of the impurity}}

The DSMC simulations described in the preceding section show that
the steady state reached by the system is in agreement with the bulk
profiles assumed in the hydrodynamic solution worked out in Sec.\
\ref{sec3}, i.e., the pressure $p_2$, the local shear rate $a$, and
the local thermal curvature $\gamma$ are practically uniform.
Moreover, the impurity properties are enslaved to those of the gas
particles, namely the system evolves to $\mathbf{u}_1=\mathbf{u}_2$,
$n_1/n_2=\text{const}$, and $T_1/T_2=\text{const}$, {in agreement
with assumptions (i)--(iii) listed below Eq.\ \eqref{III.21}}. As an
illustration, Fig.\ \ref{profiles} shows simulation data of the
pressure and velocity profiles for both the impurity and the gas
particles at $t\approx \overline{\tau}$ and $t\approx 10^3
\overline{\tau}$.

\subsection{{Thermal curvature coefficient}}
In the remainder of this section we compare the theoretical results
derived in Sec.\ \ref{sec3} for $d=3$ with the data obtained from
our DSMC simulations. Before considering properties associated with
the impurity, we first compare the shear rate dependence of the
{thermal curvature coefficient $\gamma$}. Figure \ref{gamma}
displays $\gamma$ versus $a^2$ for three values of the coefficient
of restitution $\alpha_{22}$: $\alpha_{22}=1$ (elastic case),
$\alpha_{22}=0.9$ (moderately inelastic case), and $\alpha_{22}=0.8$
(quite inelastic case). It is observed that the theory compares
 well with the simulation results for the three values of
$\alpha_{22}$ considered, even for strongly sheared gases. This
confirms the reliability of a (non-Newtonian) hydrodynamic
description for granular gases  in the bulk domain and beyond the
quasi-elastic limit, at least within the framework of the model
kinetic equations used. It is apparent from Fig.\ \ref{gamma} that,
at a given value of the reduced shear rate $a$, the value of
$\gamma$ decreases with increasing dissipation. This can be
qualitatively understood by the tendency of the collisional cooling
to produce a concave temperature profile, while the viscous heating
tends to produce a convex profile. In fact, both tendencies cancel
each other at the threshold shear rate $a_\kk$, where $\gamma=0$.
The corresponding values for $\alpha_{22}=0.9$ and
$\alpha_{22}=0.8$, are $a_\kk=0.29$ and $a_\kk=0.43$, respectively.
As noted above, our analytical solution is not mathematically well
defined for negative values of $\gamma$ (i.e., $a<a_\kk$, shaded
region in Fig.\ \ref{fig2}). This restriction obviously does not
apply to the simulations, which can reach states with $\gamma<0$.
These states also include those in the absence of shearing ($a=0$).
States with $a=0$ are interesting and some cases have been studied,
in the framework of the NS description and/or in the quasi-elastic
limit \cite{BC98}.

\subsection{{Temperature ratio}}

Let us study now the main properties characterizing the hydrodynamic
state of the impurity. The parameter space of the problem is made of
four (dimensionless) material quantities (the mass ratio
$\mu=m_1/m_2$, the size ratio $\omega=\sigma_1/\sigma_2$, and the
coefficients of restitution $\alpha_{12}$ and $\alpha_{22}$) plus
the reduced shear rate $a$. For the sake of illustration, we will
assume a common coefficient of restitution
$\alpha_{12}=\alpha_{22}=\alpha$ and a common size ($\omega=1$), so
that the parameter space becomes three-dimensional. Furthermore, we
focus on three values of $\mu$ ($\mu=2$, $\mu=1$, and $\mu=0.5$) and
three values of $\alpha$ ($\alpha=1$, $\alpha=0.9$, and
$\alpha=0.8$), so that we consider nine different systems. For each
one, we analyze the dependence of the properties of the impurity on
the shear rate. Note that, since $\omega=1$ and
$\alpha_{12}=\alpha_{22}$, in the case $\mu=1$ the impurity is
mechanically equivalent to the gas particles.

First, the breakdown of energy equipartition, as measured by the
temperature ratio $\chi=T_1/T_2$, is plotted in Fig.\ \ref{chi}. A
good agreement between theory and simulations is observed. The lack
of energy equipartition is expected because of two reasons. On the
one hand, the state of the system is far from equilibrium due to the
shearing and thus $T_1\neq T_2$ even in the elastic case
($\alpha=1$) \cite{GS93,GS03}. On the other hand, even in the
homogeneous cooling state, the inelasticity drives the system out of
equilibrium and, consequently, $T_1\neq T_2$ \cite{GD99b}. We see
from Fig.\ \ref{chi} that the impurity has a higher (lower) granular
temperature than the gas if it is heavier (lighter) than a gas
particle. This agrees with the general trend observed in experiments
\cite{WP02,FM02}. Figure \ref{chi} also shows that, for a given
value of $\alpha$, the deviation of the temperature ratio from unity
increases as the shear rate increases. Similarly, at a given value
of $a$, the deviation $\chi-1$ becomes more important with
increasing dissipation.

\subsection{{Generalized transport coefficients}}

Next, we explore the momentum and heat transport of the impurity, as
measured by the rheological quantities $\eta_1$, $N_1$, $M_1$,
$\lambda_1$, and $\phi_1$ defined by Eqs.\
\eqref{III.26}--\eqref{III.28}. Figures \ref{eta1}--\ref{M1} display
the three transport coefficients associated with the pressure
tensor. As in the case of $\chi$, the agreement between the
theoretical predictions and the simulation results is very good. It
is apparent that, regardless of the value of $\alpha$, shear
thinning effects are present, i.e., {the nonlinear shear viscosity}
$\eta_1$ decreases with increasing shear rate. Regarding the
influence of the mass ratio, we observe that, for fixed values of
$\alpha$ and $a$,  $\eta_1$ increases as the mass ratio increases.
The influence of dissipation on $\eta_1$ is smaller than that of
$\mu$. In any case, although hardly apparent in Fig.\ \ref{eta1},
the value of $\eta_1$ increases as $\alpha$ decreases at given $\mu$
and $a$. It is interesting to note that the ratio $\eta_1/\chi$ is
practically independent of $\mu$, although it exhibits a weak
dependence on $\alpha$.

The {viscometric coefficients $N_1$ and $M_1$, which measure normal
stress differences, are plotted in Figs.\ \ref{N1} and \ref{M1},
respectively}. The shearing produces a strong anisotropy in the
normal stresses: $P_{1,xx}>n_1T_1>P_{1,zz}>P_{1,yy}$. As expected,
this anisotropy increases with the shear rate. While, for given $a$
and $\alpha$, the {coefficient} $N_1$ increases as the impurity
becomes heavier, the opposite happens with the {coefficient} $M_1$.
With respect to the influence of $\alpha$, it turns out that it is
practically negligible in the case of $N_1$, while $M_1$ decreases
significantly as the system becomes more inelastic.

Finally, the two transport coefficients $\lambda_1$ and $\phi_1$
measuring the heat flux are plotted in Figs.\ \ref{lambda1} and
\ref{phi1}, {respectively}. These coefficients are quite difficult
to measure in the simulations near the threshold shear rate $a_\kk$,
since there the thermal gradient is very small. This explains the
scatter of the simulation data near $a^2=a_\kk^2$. Again, theory
compares quite well with simulations. This is rather satisfactory
especially in the case of $\phi_1$ since this {cross} coefficient
measures complex coupling effects between the velocity and
temperature gradients, which are absent in the NS regime. Figure
\ref{lambda1} shows that, analogously to what happens with $\eta_1$,
the generalized thermal conductivity $\lambda_1$ decreases with
increasing shear rate. In contrast, the cross coefficient $\phi_1$
has a non-monotonic dependence for small inelasticities. In
agreement with the behavior found for $\eta_1$ and $N_1$, both
coefficients $\lambda_1$ and $\phi_1$ decrease as the mass of the
impurity decreases, at given values of $a$ and $\alpha$. As for the
influence of $\alpha$, the results show that $\lambda_1$ and
$\phi_1$ increase with increasing dissipation, this effect being
more important for a heavy impurity than for a light impurity. We
have observed that the influence of the mass ratio on $\lambda_1$
and $\phi_1$ is significantly inhibited when one considers the
ratios  $\lambda_1/\chi^2$ and $\phi_1/\chi^2$, especially in the
former case. A remarkable counter-intuitive feature is that the
coefficient $\phi_1$ can turn out to be larger than $\lambda_1$ for
sufficiently large shear rate. This effect is more notorious as the
system becomes more inelastic and/or the impurity becomes heavier.
In fact, in the cases $\mu=1$ and $\mu=2$ with $\alpha=0.8$, one has
$\phi_1>\lambda_1$ for any shear rate larger than $a_\kk$. Taking
into account the definitions \eqref{III.28}, the situation
$\phi_1>\lambda_1$ implies that $|q_x|>|q_y|$, i.e., the shearing
induces a heat flux with a component orthogonal to the thermal
gradient that is larger than the component parallel to the gradient.

\subsection{{Preliminary DSMC results from the true Boltzmann
equation}}

{Thus far we have shown that the numerical solutions of the model
kinetic equations \eqref{MC1} with realistic boundary conditions
support the  steady-state hydrodynamic solution derived in this
paper for the same model beyond the small Knudsen number limit.
However, the important question is whether or not such a generalized
hydrodynamic description is supported by the more fundamental
Boltzmann equation.  Comparison between DSMC simulations of the
Boltzmann equation and the hydrodynamic solution of the kinetic
model shows that the answer is affirmative in the case of a
monocomponent granular gas \cite{TTMGSD01}.

When an impurity particle is embedded in the host granular gas, the
crucial point of the hydrodynamic solution worked out in section
\ref{sec3} is the enslaving of the hydrodynamic fields of the
impurity to those of the bath, as expressed by assumptions
(i)--(iii) below Eq.\ \eqref{III.21}. We have performed preliminary
DSMC simulations of the Boltzmann equation for the host gas and the
coupled Boltzmann--Lorentz equation for the impurity particle and
have observed that  the properties (i)--(iii) are indeed satisfied
in the steady state and in the bulk domain. As an illustrative
example, Fig.\ \ref{profiles2} shows the  pressure and velocity
profiles, as obtained from DSMC simulations of the Boltzmann
equation, for a system similar to that of Fig.\ \ref{profiles} but
with a larger separation  between the plates. Again, in the steady
state (and also practically during the transient regime) one has
$\mathbf{u}_1=\mathbf{u}_2$. Moreover, both $p_1$ and $p_2$ are
practically constant in the bulk region. As in Fig.\ \ref{profiles},
$p_1/\overline{n}_1>p_2/\overline{n}_2$, but this effect is smaller
than in Fig.\ \ref{profiles} because now $L$ is larger and so the
shear rate $a$ is smaller. Moreover, as exemplified by Figs.\
\ref{profiles} and \ref{profiles2}, we have observed that the
boundary effects are more important in the case of the Boltzmann
description than in that of the kinetic model. A more exhaustive
study, including the temperature ratio and the generalized transport
coefficients, is ongoing and will be published elsewhere
\cite{VSG08}.}

\section{Conclusions\label{sec6}}
In this paper we have analyzed the transport properties of
impurities immersed in a granular gas under nonlinear steady planar
Couette flow. We have focused {on} situations where the shear rate
is large enough as to make the viscous heating term prevail over the
inelastic cooling term in the energy balance equation. In these
conditions the NS description is in general inadequate, as
illustrated by Fig.\ \ref{fig2}, and so a more fundamental kinetic
theory is needed. Due to the mathematical complexity of the
Boltzmann equation, here we have used a kinetic model  for granular
mixtures recently proposed by the authors \cite{VGS07}. Our approach
differs from a recent work \cite{LMG07} on a bidisperse granular
fluid under Couette flow, where a continuum description is used. In
addition, the present work extends to inelastic collisions a
previous study \cite{GS93} carried out for ordinary gaseous
mixtures.

Two different and complementary routes have been considered. First,
an exact \emph{hydrodynamic} (or ``normal'') solution for the steady
state has been found. This solution applies for arbitrarily large
shear rates $a$ (larger than the threshold value $a_\text{th}$
corresponding to the simple shear flow) and arbitrary values of the
parameters of the system (coefficients of restitution, masses, and
sizes). Progress has been made taking advantage of a formal mapping
between the kinetic equation for the gas particles (whose exact
hydrodynamic solution was found in Ref.\ \cite{TTMGSD01}) and the
kinetic equation for the impurity. This formal mapping is possible
once it is guessed that the hydrodynamic profiles of the impurity
are enslaved to those of the gas particles, i.e., $n_1/n_2$ and
$T_1/T_2$ are uniform and $\mathbf{u}_1=\mathbf{u}_2$ (no
diffusion).  {Second, we have solved the set of two coupled kinetic
equations by means of a DSMC method \cite{Bird}, with realistic
boundary conditions. The numerical solution shows, in the context of
our kinetic model description, the validity of the assumptions we
make in the calculation of the theoretical solution. Furthermore, we
have not found ranges of parameter values where these assumptions
are not accurately fulfilled in the bulk of the fluid. Thus, an
important corollary of this work is that under Couette flow and for
our kinetic model the impurity never shows {steady-state} diffusion
with respect to the granular gas where it is immersed (even in a
strongly sheared system).}

In order to characterize the nonequilibrium state of the impurity,
we have selected a number of relevant dimensionless coefficients.
The basic one is the temperature ratio $\chi=T_1/T_2$, quantifying
the lack of energy equipartition between both species. The momentum
flux defines three independent coefficients: the nonlinear shear
viscosity $\eta_1$, Eq.\ \eqref{III.26}, and the two {viscometric
coefficients (or normal stress differences)} $N_1$ and $M_1$, Eq.\
\eqref{III.27}. Similarly, the heat flux defines the nonlinear
thermal conductivity $\lambda_1$ and the cross coefficient $\phi_1$,
Eq.\ \eqref{III.28}. Notice that the coefficients $N_1$, $M_1$, and
$\phi_1$ do not have counterparts at the NS level. In particular,
the coefficient $\phi_1$ is interesting because it accounts for a
component of the heat flux orthogonal to the thermal gradient,
induced by the shearing.

Comparison between the exact solution and the DSMC simulations shows
a good agreement, thus indicating the existence of a hydrodynamic or
normal solution, even under extreme conditions, beyond the NS
regime. The results show that, in general,  $T_1$ is higher (lower)
than $T_2$ if $m_1$ is larger (smaller) than $m_2$. Moreover, as
expected, the deviation of the temperature ratio $\chi$ from unity
increases as the inelasticity and/or the shear rate increase.
Concerning the generalized coefficients $\eta_1$ and $\lambda_1$, it
is observed that both decrease as the shear rate increases, while
they increase with increasing dissipation and mass ratio $m_1/m_2$.
As expected, the anisotropy of the normal stresses increases as the
shear rate increases. In addition, as the impurity becomes heavier,
the difference between the $xx$ and $yy$ stresses increase, while
the difference between the $zz$ and $yy$ stresses decrease. The
latter effect is also present when the system becomes more
inelastic. Finally, in general, the cross coefficient $\phi_1$ does
not present a monotonic dependence on the shear rate. However, like
in the cases of $\eta_1$ and $\lambda_1$, the coefficient $\phi_1$
increases as the mass of the impurity and/or dissipation increase.
Interestingly, the latter effect is so remarkable that $\phi_1$ can
be larger than $\lambda_1$ (and hence $|q_x|>|q_y|$) if the impurity
is sufficiently massive or the system is sufficiently inelastic.

The work carried out in this paper can be extended along several
lines. On the one hand, since the states considered here have been
restricted to conditions where $\gamma>0$ ($a>a_\text{th}$), it
would be desirable to extend the analysis to the complementary
situations where $\gamma<0$ ($a<a_\text{th}$, shaded region in Fig.\
\ref{fig2}). While the simulation method does not present any
technical difficulty in the latter case, the analytical solution
found in this paper involves $\sqrt{\gamma}$ [see, for instance,
Eqs.\ \eqref{III.9}--\eqref{III.14}] and so is not mathematically
well defined when $\gamma<0$. {However, we have observed that an
analytical continuation of the solution accounts well for the
simulation results for a range of negative values of $\gamma$
\cite{VGS08}. Another} possible alternative to overcome this
technical difficulty is to carry out a perturbation solution in
powers of $\gamma$, exploiting the fact that $|\gamma|$ is a small
parameter in the region $a<a_\text{th}$, as preliminary computer
simulation results show. A second interesting problem is the
extension of the tracer limit results derived here to a general
bidisperse mixture with arbitrary composition. The main idea would
be to guess hydrodynamic profiles for the mixture similar to those
of a monodisperse system \cite{TTMGSD01}, along with a common flow
velocity and uniform mole fractions and temperature ratios. Finally,
the theoretical results predicted by the kinetic model will be
confronted with those obtained by DSMC simulations of the true
Boltzmann equation. {Our preliminary results show that the good
agreement found in the monodisperse case \cite{TTMGSD01} extends to
the case of mixtures, at least at a semi-quantitative level}.

\acknowledgments

This research  has been supported by the Ministerio de Educaci\'on y
Ciencia (Spain) through Programa Juan de la Cierva (F.V.R.) and
Grant No.\ {FIS2007-60977}, partially financed by FEDER funds.

\appendix*
\section{Transport properties associated with the impurity
{at the threshold shear rate}\label{appA}} In this Appendix we
derive the explicit expressions for the transport coefficients of
the impurity along the threshold shear rate $a_\kk(\alpha_{22} )$.
They are obtained by taking the limit $\gamma \rightarrow 0^{+}$ in
the corresponding expressions of Sec.\ \ref{sec3}. A similar study
was carried out in Ref.\ \cite{G02} by applying Grad's method to the
Boltzmann equation.

First, note that when $y\to 0^+$ the function $\theta(w,y,z)$
defined by Eq.\ \eqref{III.14} goes to infinity, so that one can
make use of the asymptotic expansion of the complementary error
function \cite{AS72}, i.e., \beq \sqrt{\pi}\theta
e^{\theta^2}\text{erfc}(\theta)\approx 1-\frac{1}{2\theta^2},\quad
\theta\gg 1. \label{A1} \eeq Inserting this expansion into Eq.\
\eqref{III.12} and performing the integral, one obtains \beq
F_{0,m}(y,z)\approx -
\frac{4m!}{z^2}\left[(1+z)^{-(1+m)}+(1+2z)^{-(1+m)}-2^{2+m}(2+3z)^{-(1+m)}\right]y,\quad
y\ll 1. \label{A2} \eeq {Since $F_{1,m}(y,z)=y\partial
F_{0,m}(y,z)/\partial y$, it follows that $F_{1,m}(y,z)\approx
F_{0,m}(y,z)$} to first order in $y$. Furthermore, the functions
$X(\theta)$ and $Y(\theta)$ defined by Eqs.\ \eqref{X} and
\eqref{Y}, respectively, behave as \beq X(\theta)\approx
\frac{1}{\theta},\quad Y(\theta)\approx \frac{3}{2\theta},\quad
\theta\gg 1. \label{A3} \eeq Therefore, \beq G(y,z)\approx
\frac{d+4}{z}\left[-(1+2z)^{-2}+4(2+3z)^{-2}\right]\sqrt{2y},\quad
y\ll 1, \label{A4} \eeq \beq H(y,z)\approx
\frac{18}{z}\left[-(1+2z)^{-4}+16(2+3z)^{-4}\right]\sqrt{2y},\quad
y\ll 1. \label{A5} \eeq

Since {both $F_{0,m}(\widetilde{\gamma},\zetat)$ and
$F_{1,m}(\widetilde{\gamma},\zetat)$ go to zero} when $\gamma\to 0$,
Eq.\ \eqref{n9} becomes
\begin{equation}
\label{A6}
d\left(\frac{T_1}{T_{12}}-\frac{1}{1+\zetat}\right)-\frac{2\widetilde{a}_\kk^2}
{(1+\zetat)^3}=0.
\end{equation}
This is a fourth-degree  algebraic equation whose physical solution
gives the temperature ratio $\chi$ in the simple shear flow. Once
$\chi$ is known, the transport coefficients are readily obtained.
The coefficients associated with the momentum transport are, from
Eqs.\ \eqref{III.29}--\eqref{III.31},
\begin{equation}
\label{A7}
\eta_1=\frac{T_{12}}{T_1}\frac{\chi}{(1+\zetat)^2},
\end{equation}
\begin{equation}
\label{A8}
N_1=2\frac{T_{12}}{T_1}\frac{\widetilde{a}_\kk^2}{(1+\zetat)^3} ,
\quad M_1=0.
\end{equation}

The evaluation of the generalized thermal conductivity $\lambda_1$
at $\gamma=0$  from Eq.\ \eqref{III.32} is trickier than before
since substitution of Eq.\ \eqref{A7} into \eqref{III.32} yields an
indeterminate result. This difficulty is circumvented by first
eliminating $\widetilde{a}^2$ between Eqs.\ \eqref{n9} and
\eqref{III.32} and replacing $\eta_1$ by its expression
\eqref{III.29}. The result expresses $\lambda_1$ in terms of the
functions {$F_{0,m}(\widetilde{\gamma},\zetat)$ and
$F_{1,m}(\widetilde{\gamma},\zetat)$}. Then, the asymptotic value
\eqref{A2} is used and the limit $\widetilde{\gamma}\to 0$ is taken.
The final result is \beq
\lambda_1=\left(\frac{T_{12}}{T_1}\right)^2\frac{2\chi^2}{2+7\zetat+6\zetat^2}
\left[1+\frac{6}{d+2}\frac{12+42\zetat+37\zetat^2}
{(2+7\zetat+6\zetat^2)^2}\widetilde{a}_\kk^2\right]. \label{A9} \eeq
The limit $\gamma\to 0$ of the cross coefficient $\phi_1$ is easily
obtained from Eq.\ \eqref{III.33} as \beq
\phi_1=\frac{2}{d+2}\left(\frac{T_{12}}{T_1}\right)^2\chi^2\frac{4+7\zetat}{(2+7\zetat+6\zetat^2)^2}\widetilde{a}_\kk
\left[d+4+18\frac{8+28\zetat+25\zetat^2}
{(2+7\zetat+6\zetat^2)^2}\widetilde{a}_\kk^2\right], \label{A10}
\eeq where use has been made of Eqs.\ \eqref{A4} and \eqref{A5}.
Despite the fact that there is no heat flux in the simple shear
flow, Eqs.\ \eqref{A9} and \eqref{A10} are intrinsic transport
coefficients characterizing the state of the system. {Equations
\eqref{A7}--\eqref{A10} also describe the transport properties of
the Couette flow with a temperature profile linear in $s$.}

Equations \eqref{A7}--\eqref{A10}, when particularized to an
impurity mechanically equivalent to the particles of the host gas,
are consistent \cite{note} with the results reported in Appendix D
of Ref.\ \cite{TTMGSD01}.


\begin{thebibliography}{99}
\bibitem{note1}
Note that, for convenience, the expressions for $\gamma$, the
pressure tensor, and the heat flux have been written in a
representation slightly different from that of  Ref.\
\cite{TTMGSD01}. Of course, both representations are completely
equivalent.
\bibitem{note}
Note, however, that Eq.\ (D4) of Ref.\ \protect\cite{TTMGSD01} is
restricted to a three-dimensional gas.
\end{thebibliography}
\end{document}